\documentclass[twocolumn,showpacs]{revtex4-1}
\usepackage{graphicx}
\usepackage{color}
\graphicspath{{pict/}{}}

\newcounter{Fig}

\newcommand{\be}{\begin{equation}}
\newcommand{\ee}{\end{equation}}

\begin{document}
\title{Surface solitons in quasiperiodic nonlinear photonic lattices}
\author{Alejandro J. Mart\'inez and Mario I. Molina}
\affiliation{Departamento de F\'isica, Facultad de Ciencias,
Universidad de Chile, Santiago, Chile and Center for Optics and
Photonics (CEFOP), Casilla 4016, Concepci\'on, Chile.}
\pacs{42.25.Dd, 42.65.Tg, 42.65.Sf, 42.65.wi}
\begin{abstract}
We study discrete surface solitons in semi-infinite, one-dimensional, nonlinear (Kerr), quasiperiodic  waveguide arrays of the Fibonacci and Aubry-Andr\'e types, and explore different families of localized surface modes, as a function of optical power content (`nonlinearity') and quasiperiodic strength (`disorder'). We
find a strong asymmetry in the power content of the mode as a function of the propagation constant, between the cases of focussing and defocussing nonlinearity, in both models.  We also examine the dynamical evolution of a completely-localized initial excitation at the array surface.  We find that in general, for a given optical power, a smaller quasiperiodic strength is required to effect localization at the surface than in the bulk. Also, for fixed quasiperiodic strength, a smaller optical power is needed to localize the excitation at the edge than inside the bulk.
\end{abstract}

\maketitle

\section{INTRODUCTION}
Optical waveguide arrays and photonic lattices have proven particularly 
useful to the study and observation of  
the general phenomenology of wave localization arising 
from the interplay of nonlinearity, periodicity (or lack of it), 
disorder, boundary effects, geometry and dimensionality. Among
other phenomena, it has permitted the direct observation of 
Anderson localization~\cite{AL-exp}, 
Tamm states in truncated optical lattices~\cite{tamm} and  boundary
effects on Anderson localization~\cite{Szameit-anderson}. 
Regarding the effects of quasiperiodicity in photonic systems,
low-loss resonant modes have been predicted in aperiodic 
nanopillar waveguides~\cite{nanopillars}, and localization of 
interface modes at the boundary between metal and Fibonacci 
quasiperiodic structure have been recently studied~\cite{metal-fibonacci}. 
Also, for a one-dimensional quasiperiodic waveguide array 
known as the Aubry-Andr\'{e} (AA) model, a localization transition 
was experimentally observed~\cite{Morandotti-AA} 
in agreement with the earlier prediction~\cite{AA}. 
Moreover, and in contrast to periodic systems, quasiperiodic lattices support the existence of linear localized surface modes ~\cite{faso} and a rich 
hierarchical band-gap structure~\cite{niu}. For this reason the region of existence (in parameter space) of surface gap solitons is expected to be larger in the quasiperiodic
case than in the periodic case.

Concerning the interplay of boundary and disorder, 
Szameit et. al.\cite{Szameit-anderson} observed experimentally that, in a 
truncated disordered waveguide array, localization due to 
disorder is weakened near the boundary of the truncated array. 
This result, in addition to the well-known fact 
that nonlinear localization at the edge of a 
truncated periodic nonlinear (Kerr) lattice is also weakened by 
the presence of a boundary~\cite{Molina-Surface}, reinforces the 
idea that in one dimension, the boundaries are ``repulsive''. 
The whole story is not that simple, however. A recent work\cite{MLT} extended Szameit's studies by exploring the interplay of boundaries, disorder and nonlinearity, finding that the boundary is ``repulsive'' or not depending on the relative strength of nonlinearity and disorder: For weak nonlinearity and moderate disorder, localization is weaker at the surface than in the bulk, in agreement with Szameit's results. However, for relatively strong disorder 
and/or nonlinearity it is as easy to localize an excitation at the edge as in the bulk.

In this work we explore another side of these issues by replacing in these studies the presence of uncorrelated disorder by the much weaker loss of periodicity, produced by quasiperiodicity. That is, we examine the interplay of boundary effects, quasiperiodicity and  nonlinearity, in 
two different semi-infinite nonlinear optical waveguide arrays. 
In one of them, the linear index of refraction follows 
the Fibonacci sequence, while in the other one, the index of 
each fiber varies in space in an  incommensurate manner, according to the AA model. 
The strength of quasiperiodicity is varied by adjusting the relative refractive 
index contrast of the guides, while nonlinearity is regulated by the power content
of the input beam.

We find different families of discrete solitons with different power thresholds: 
High power solitons associated to total internal reflection (TIR) gaps 
with finite power threshold, and low power solitons
associated to Bragg reflection (BR) gaps where the threshold value
depend on the symmetry of linear modes and their propagation constants.
Contrary to what happens in periodic systems, here we have localized linear modes and
extended asymmetric linear modes that can originate discrete solitons with
zero power threshold. For a fixed  distribution of refractive indices, we find a strong asymmetry between the focussing and defocussing cases, when the quasiperiodicity of the array is diagonal, i.e., when it affects the index of refraction of the guides but not their mutual coupling.  We also examine the dynamical evolution of a completely-localized initial excitation at the array surface.  We find that in general, for a given power content, a smaller quasiperiodic strength is required to effect localization at the surface than in the bulk. Also, for a fixed quasiperiodic strength, a smaller power is needed to localize the excitation at the edge than inside the bulk. This behavior is in marked contrast to the case of
uncorrelated disorder and weakens the usual assumption that a surface acts in 
a `repulsive' manner\cite{Molina-Surface,Szameit-anderson}.

This paper is organized as follows. In Sec. II we describe the two kinds of 
quasiperiodicity (substitutional and incommensurate arrays) and we discuss the
appropriate adimensionalization in each case. In Sec. III we study the different
families of surface localized modes and their linear stability. In Section IV
we examine the interplay between quasiperiodicity,
boundary and nonlinear effects on the dynamical evolution of an
initially localized input beam and we compared the results obtained for
surface and bulk excitation. In section V we discuss briefly the
nondiagonal quasiperiodic case and finally, section VI concludes the paper.

\section{MODEL}

Let us consider a semi-infinite, weakly-coupled, one-dimensional quasiperiodic 
array of  nonidentical nonlinear optical single-mode waveguides. In the coupled-mode framework,
the electric field $E(x)$ propagating along the array  
can be presented as a superposition of the modes of each waveguide, 
$E(x) = \sum_{n} E_{n} \phi_n(x)$, where $E_{n}$ is the amplitude of the 
(single) guide mode $\phi_n(x)$ centered on the $n$th site. The evolution equations for the modal amplitudes $E_{n}$ take the form 
\be
\left(i\frac{d}{dz}+\epsilon_n\right)E_n + V(E_{n+1}+E_{n-1}) +\gamma|E_n|^2E_n
= 0
\label{DNLS}
\ee
where $n$ denotes the position of a guide center,
$z$ is the longitudinal propagation coordinate, $\gamma$ is the
effective nonlinear coefficient, $V$ is the coupling between
nearest-neighbor guides and $\epsilon_n$ is the refractive index
of the $n$-th guide. Equation (\ref{DNLS}) is known as the 
Discrete Nonlinear Schr\"{o}dinger (DNLS) equation and is also
found in the study of intrinsic localized modes~\cite{ILM}, 
molecular crystals~\cite{molecular crystals},
biopolymers~\cite{biopolymers}, arrays of Josephson junctions~\cite{junctions}, 
Bose-Einstein condensates in magneto-optical traps~\cite{BE}, among others.

In this work the sequence 
$\{\epsilon_{n}\}$ will not be random but aperiodic.
We will consider two kinds of quasiperiodic arrays: substitutional 
(Fibonacci) and incommensurate (AA).

\begin{figure}[t]
\includegraphics[width=8cm]{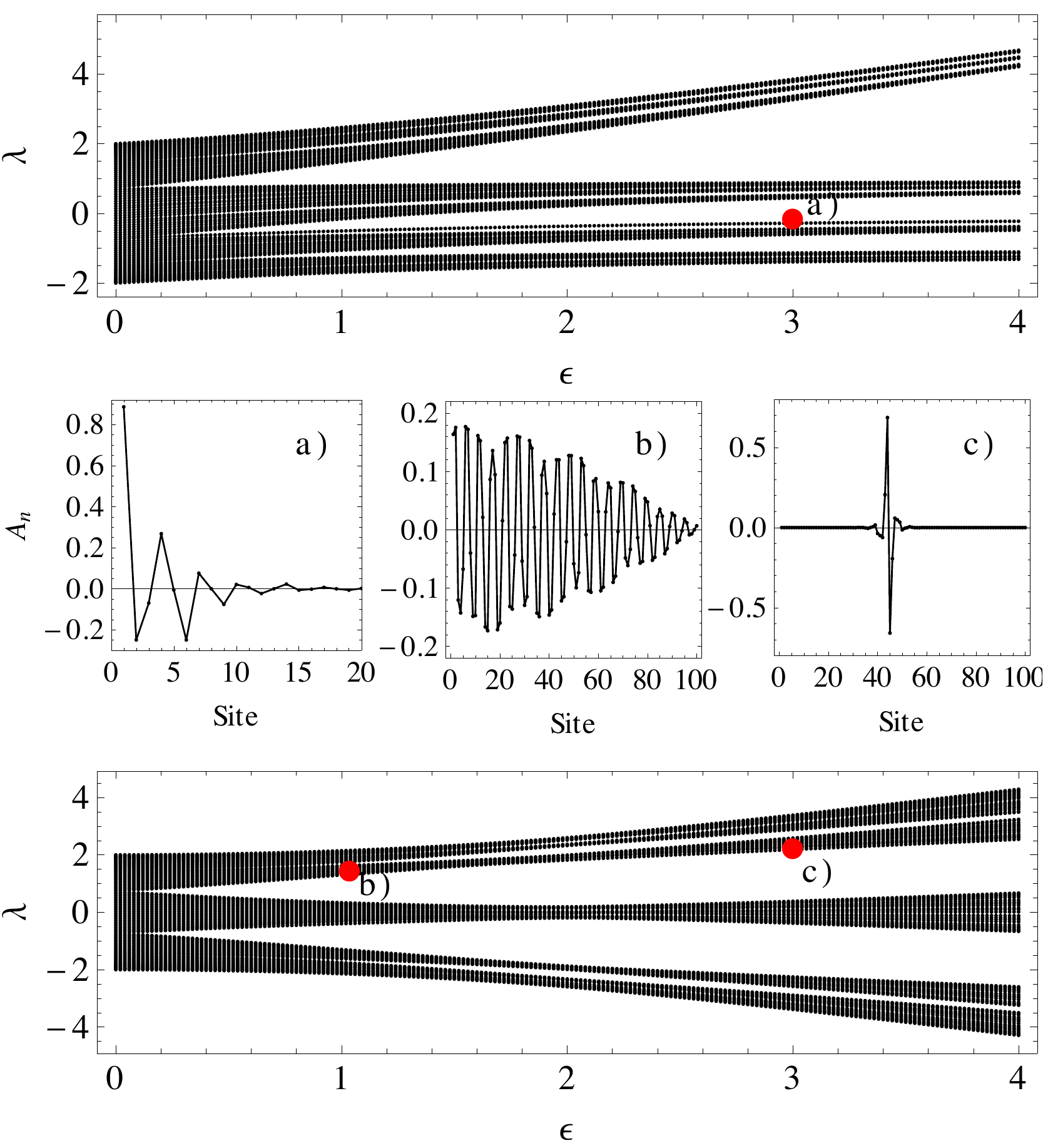}
\caption{(Color online). Linear ($\gamma=0$) spectrum as a function of $\epsilon$
for Fibonacci (top) and Aubry-Andr\'e (bottom) arrays.
For the AA model, $\chi = (\sqrt{5}+1)/2$. (a) Fibonacci surface mode. 
(b) and (c) show an asymmetric extended mode for the AA model, before ($\epsilon=1$) and after ($\epsilon=3$) the
localization transition (at $\epsilon=2$), respectively, and for the same eigenvalue.}
\label{fig1}
\end{figure}

\subsection{Substitutional arrays}
The first case corresponds to a substitutional array formed by the successive application of the Fibonacci substitutional rule: 
$\{\epsilon_a\rightarrow \epsilon_a\ \epsilon_b$, $\epsilon_b\rightarrow \epsilon_a\}$, starting from  $\epsilon_0=\epsilon_a$. This leads to an aperiodic sequence of
propagation constants: $\epsilon_a\ \epsilon_b\ \epsilon_a\ \epsilon_a\ \epsilon_b\ 
\epsilon_a\ \epsilon_b{\ldots}$, where the first value corresponds to the value at the boundary
of our semi-infinite array. We can redefine $\epsilon_a$ and
$\epsilon_b$ using an appropriate adimensionalization in term of a
single parameter $\epsilon$ (see Sec.II.C.).
In practice, since our array is of finite length $N\gg 1$, we 
only need to apply the substitutional rule a finite number of times. In the absence of nonlinearity ($\gamma=0$),
the linear modes are described by $A_n(z)=A_ne^{i\lambda z}$, reducing
Eq.(\ref{DNLS}) to an eigenvalues problem. In the limit $N\rightarrow \infty$, the linear spectrum converges 
to a fractal structure, with two main bands (Fig.1), each of which is in turn endowed with 
an internal, self-similar structure of bands and gaps~\cite{basics}.
The eigenstates has a self-similar structure too in accordance with
the aperiodicity of the array.
Among these states, there are an
infinite number of surface localized modes. For a finite array, and depending on the value 
of $N$, the number of such linear surface states is finite, appearing as isolated levels 
of the linear spectrum. It is not easy to determine how many of these linear surface states 
exist for a given $N$. For instance, for an array with $N=89$ waveguides, we find at least 
one  surface localized mode (Figs. 1a,2b).

\begin{figure}[t]
\includegraphics[width=2cm]{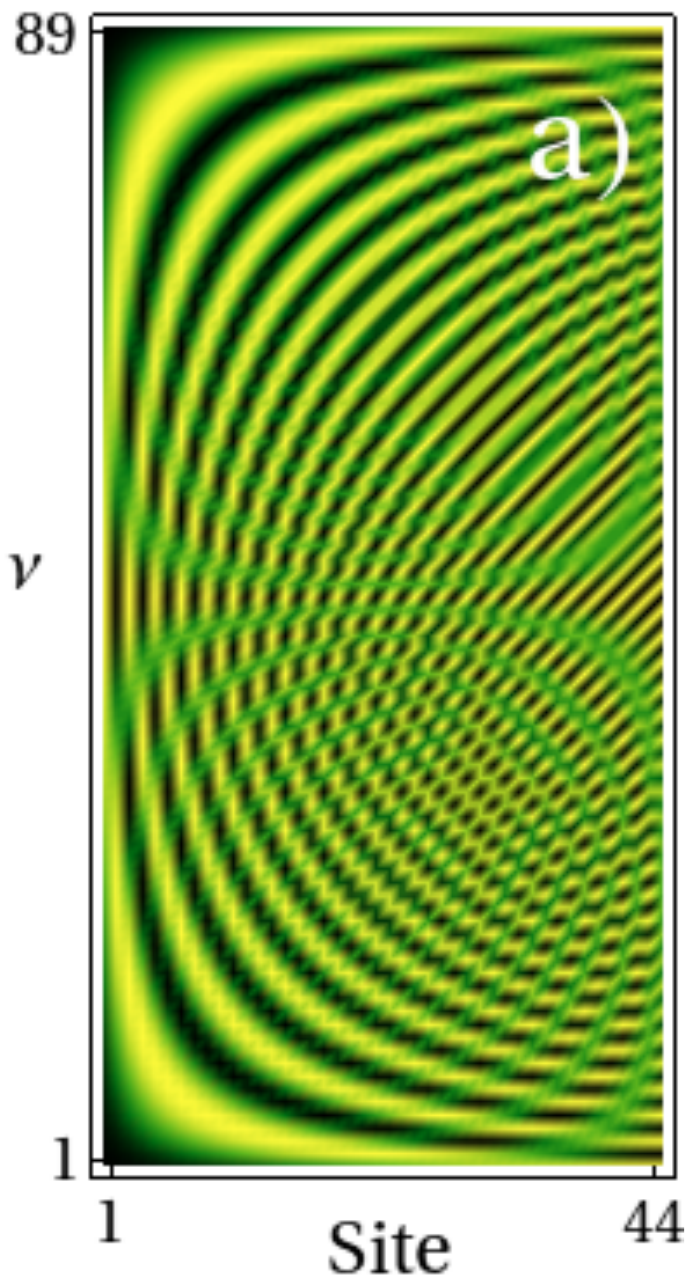}
\includegraphics[width=2cm]{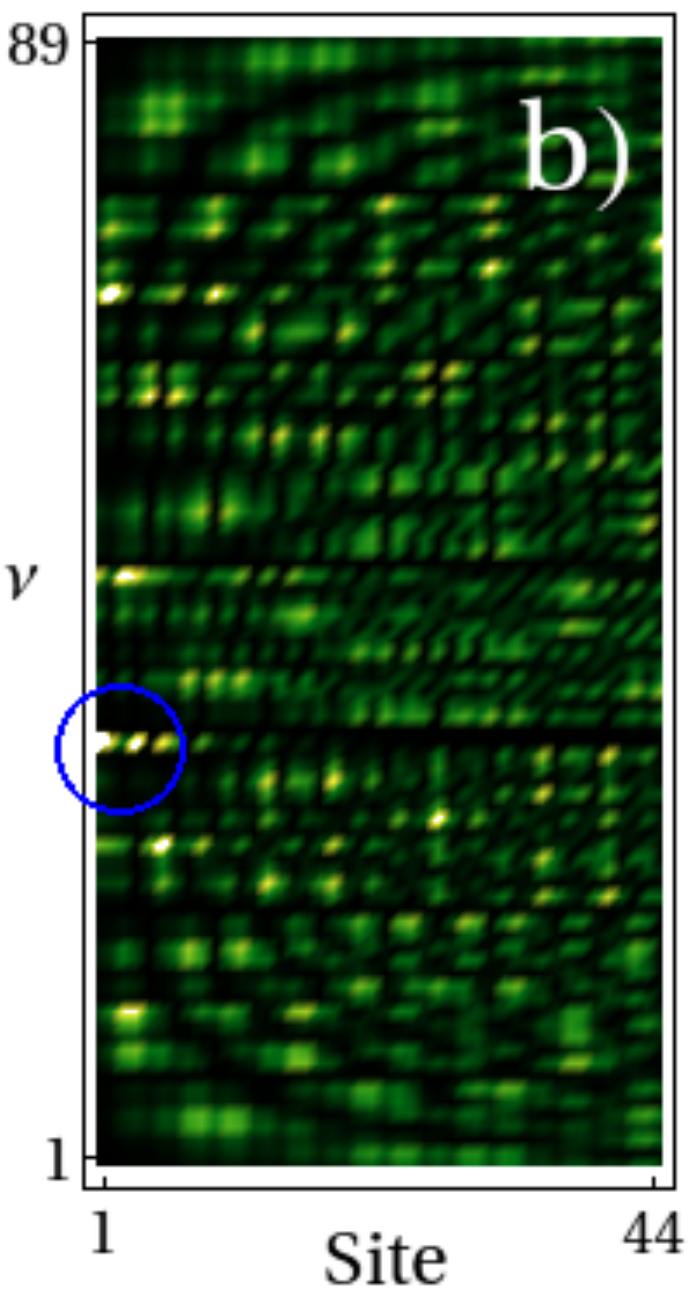}
\includegraphics[width=2cm]{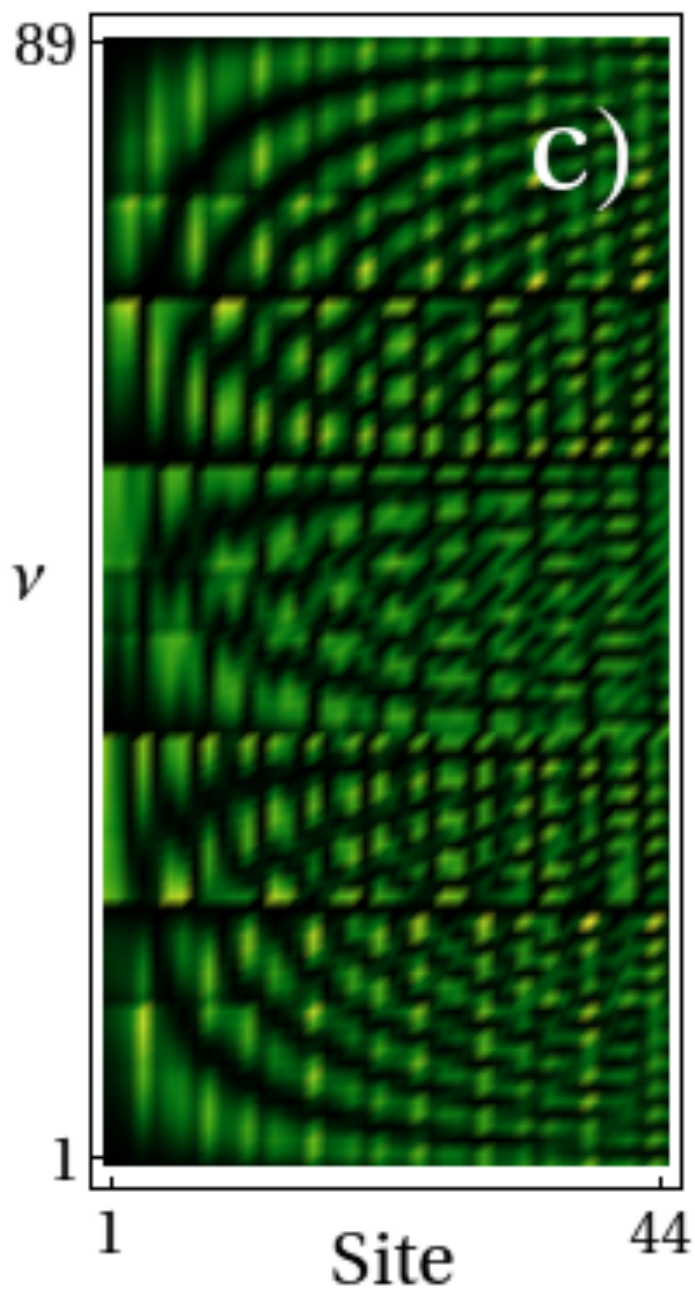}
\includegraphics[width=2cm]{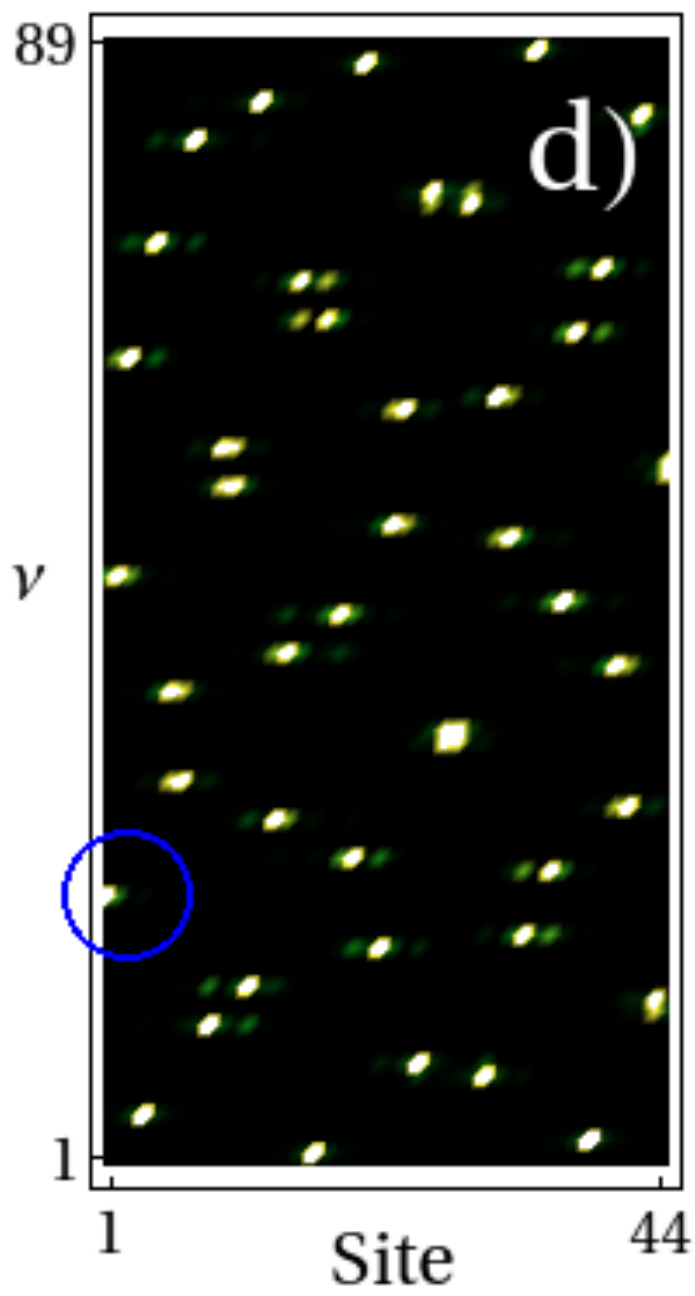}
\caption{(Color online). Intensity distribution of the linear modes for (a) periodic array,
i.e., $\epsilon=0$ in both models, (b) Fibonacci array for 
$\epsilon=1$, (c) and (d) refer to the AA array before ($\epsilon=1$) and after ($\epsilon=3$) the localization transition, respectively. The blue circle marks linear
surface states. Only the half of each mode is shown to aid the visualization of 
the surface modes. Parameter $\nu$ labels the eigenstates
in increasing order of eigenvalue (mode propagation constant).}
\label{fig0}
\end{figure}

\subsection{Incommensurately modulated arrays}

Another way to generate a linear quasiperiodic array is by means of a 
spatial modulation of the refractive index that is incommensurate 
with the natural period of the array. An interesting example of this 
is the Aubry-Andr\'{e} (AA) model~\cite{AA}, where the propagation constant of each 
waveguide is given by 
\be
\epsilon_n =\epsilon_0 +\eta\cos(2\pi n \chi),
\ee
where $\epsilon_{0}$ is the background value, $\eta$ is the strength of the modulation, and $\chi$ is the ratio between the period of the modulation and the natural lattice period.
Thus, the array is incommensurate when $\chi$ is an irrational number. The linear ($\gamma=0$) AA model 
displays a sharp localization transition at $\left|\eta/V\right|=2$:
For $\left|\eta/V\right|<2$, all eigenstates are extended, while for
$\left|\eta/V\right|>2$, they are all localized (see Fig.2). 
Moreover, in the
localized regime, the eigenvalue spectrum is composed by isolated
points forming a fractal structure such that its complement is a dense
set~\cite{AA_theory}.

\subsection{Adimensionalization}

For simplicity, we render Eq.(\ref{DNLS}) dimensionless by defining 
$E_n(z) = \sqrt{V/\gamma}A_n(\xi)e^{i\phi\xi}$, with $\xi = Vz$. 
Equation (\ref{DNLS}) reads now
\be
i\frac{d}{d\xi}A_n + (A_{n+1}+A_{n-1})+\delta_n A_n
+\mbox{sgn}(\gamma)|A_n|^2A_n
= 0\label{eq:2}
\ee
where $\delta_n = (\epsilon_n/V)-\phi$ and positive (negative) $\gamma$ denotes the focussing (defocussing) nonlinearity case. By a judicious
choice of $\phi$, one can describe both, The Fibonacci model and the
AA model. For the Fibonnaci model we define 
$\phi$ as  $\epsilon_a/V$ obtaining $\delta_n = 0$ or 
$\epsilon$ and, for AA model we define $\phi = \epsilon_0/V$ 
obtaining $\delta_n = \epsilon\cos(2\pi n \chi)$, with
$\epsilon=(\epsilon_b-\epsilon_a)/V$ and $\epsilon = \eta/V$,
as the relevant control parameters for the Fibonacci and AA models, 
respectively. The value of these parameters measure the strength of the
quasiperiodic potential.

The power content $P = \sum_{n} |A_{n}|^{2}$, is a conserved 
quantity in both models, and useful to characterize the different families 
of nonlinear modes.

\begin{figure}[t]
\includegraphics[height=3.3cm]{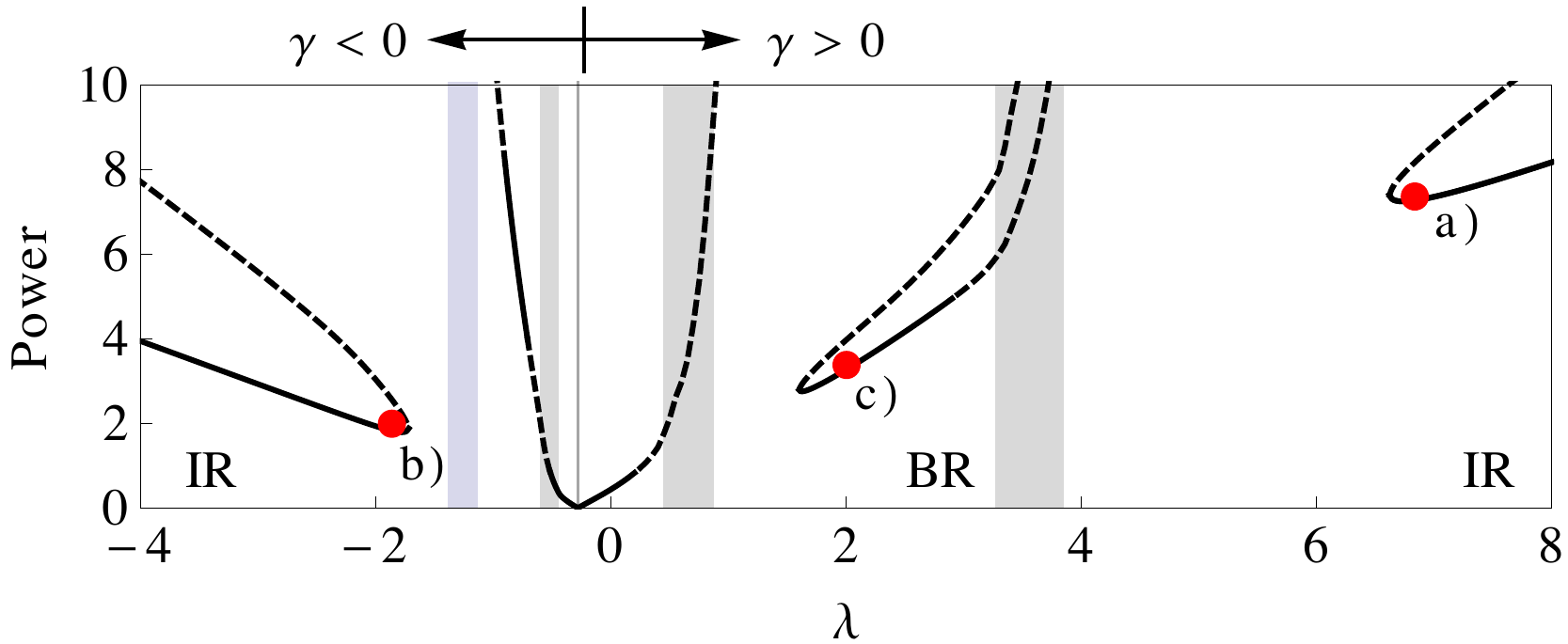}
\includegraphics[height=3.3cm]{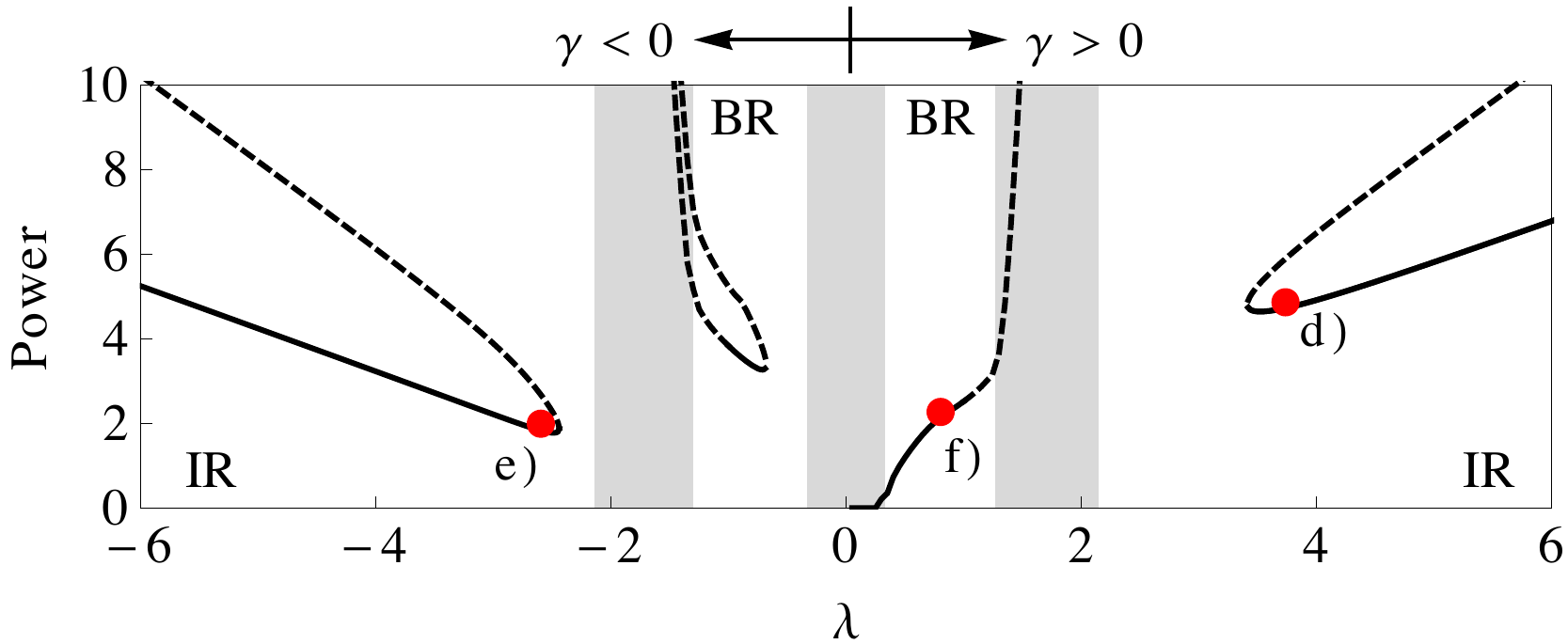}
\caption{(Color online). Power vs propagation constant for different families of 
nonlinear surface modes for both focussing and defocussing nonlinearities,
and $\epsilon>0$. Curves with positive (negative) $\lambda$ correspond to the focussing (defocussing) regime. Solid (dashed) curves denote
stable (unstable) branches. Top: Fibonacci array for $\epsilon=3$. Bottom:
AA array for $\epsilon = 1$ (phase of linear extended states).}
\label{fig2}
\end{figure}

\section{SURFACE SOLITONS}
We look for stationary localized surface mode solutions of (\ref{eq:2}). 
We pose the  {\em ansatz} $A_n(z) = A_n e^{i\lambda z}$, obtaining the coupled
system of algebraic equations
\be
-\lambda A_n + (A_{n+1}+A_{n-1})+\delta_n A_n+\mbox{sgn}(\gamma)|A_n|^2A_n
= 0,\label{eq:3}
\ee
where $A_{n}$ is real. Equation (\ref{eq:3}) possesses the useful staggered-unstaggered symmetry $(\lambda,\gamma,\epsilon,A_n)\rightarrow(-\lambda,-\gamma,-\epsilon,(-1)^n
A_n)$. This means that only a half of the $\lambda-\epsilon$ space needs to be explored.
The regions in parameter space where the nonlinear modes are
equivalent are: $\{\gamma>0,\epsilon>0\}\cup \{\gamma<0,\epsilon<0\}$
which we denote as Type I nonlinear modes, and
$\{\gamma>0,\epsilon<0\}\cup \{\gamma<0,\epsilon>0\}$, denoted as Type II
nonlinear modes. For each type, regions with $\gamma>0$ and $\lambda>0$ are associated to unstaggered modes, while regions with $\gamma<0$ and $\lambda<0$ are associated to
staggered modes. In the remaining of this section, we will fix $\epsilon>0$ and 
both signs of $\gamma$ to examine the nonlinear surface modes present in the system. 
Equations (\ref{eq:3}) are solved by a
straightforward implementation of a multi-dimensional Newton-raphson
method~\cite{Molina-Surface}, using the uncoupled (anticontinuum) limit to 
start the iteration process . The linear stability of
each mode is examined in the usual manner, by introducing a 
perturbed solution of the
kind  $A_n(\xi) = \left(A_n + \psi_n(\xi) \right)e^{i\lambda \xi}$, with $|\psi_n|\ll 1$, where
$\psi_n$ is the (complex) perturbation. After replacing in Eq.(\ref{eq:2}) and after linearizing 
in $\psi_{n}$, we obtain the evolution equation for the linear perturbation:
\be
i\frac{d}{d\xi}\psi_n=\left(\lambda-\delta_n\right)\psi_n-(\psi_{n+1}+\psi_{n-1})-\mbox{sgn}(\gamma)A_n^{2}(2\psi_n+\psi_n^*).
\ee
Next, the eigenvalue spectrum for perturbations of the type $\psi_{n}(\xi)=\psi_{n} \exp(i \Omega \xi)$ is examined, and the stability (instability) of the mode is inferred from the absence (presence) of complex values of the perturbation eigenvalues $\{\Omega\}$.

\begin{figure}[t]
\includegraphics[width=3.5cm]{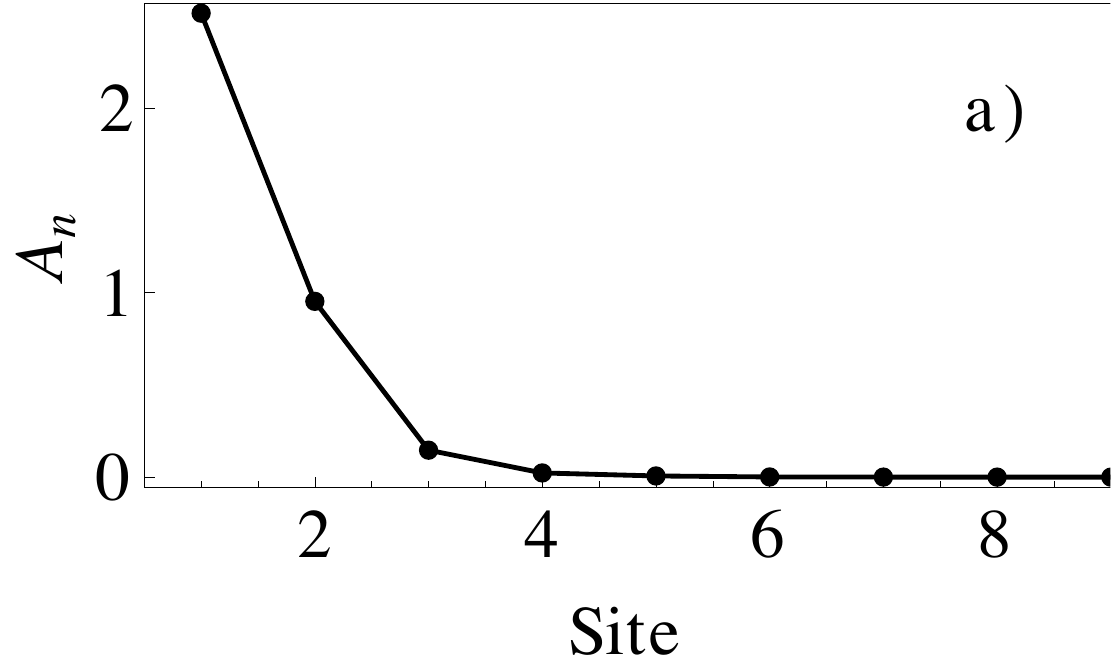}
\includegraphics[width=3.5cm]{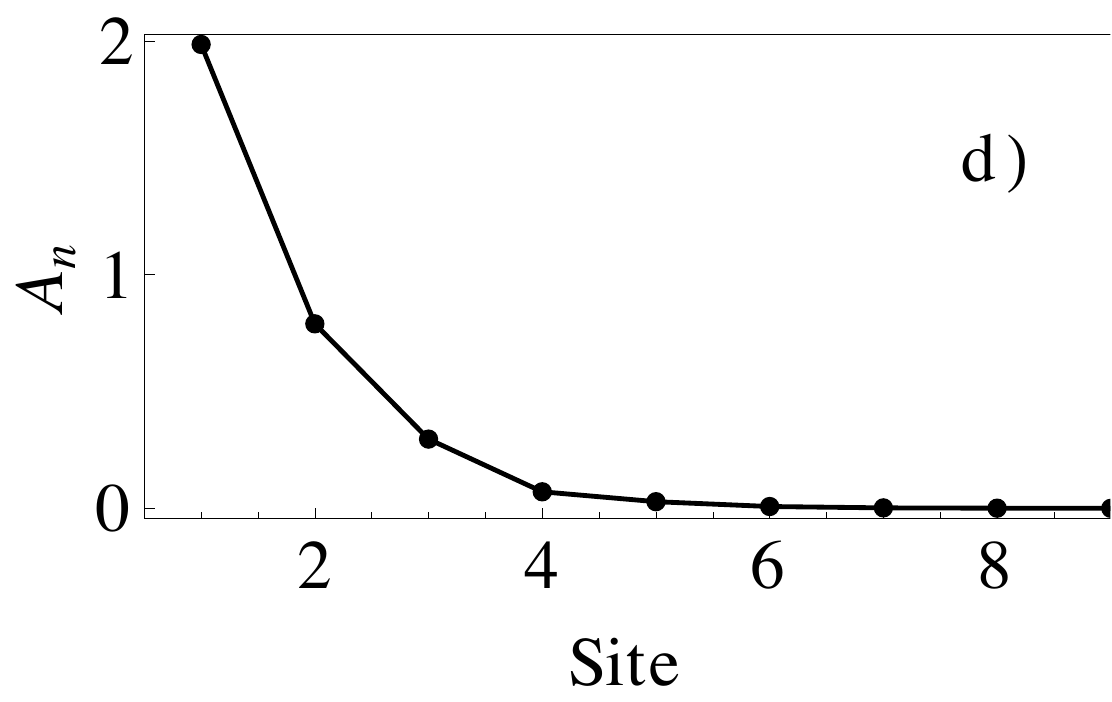}\\
\includegraphics[width=3.5cm]{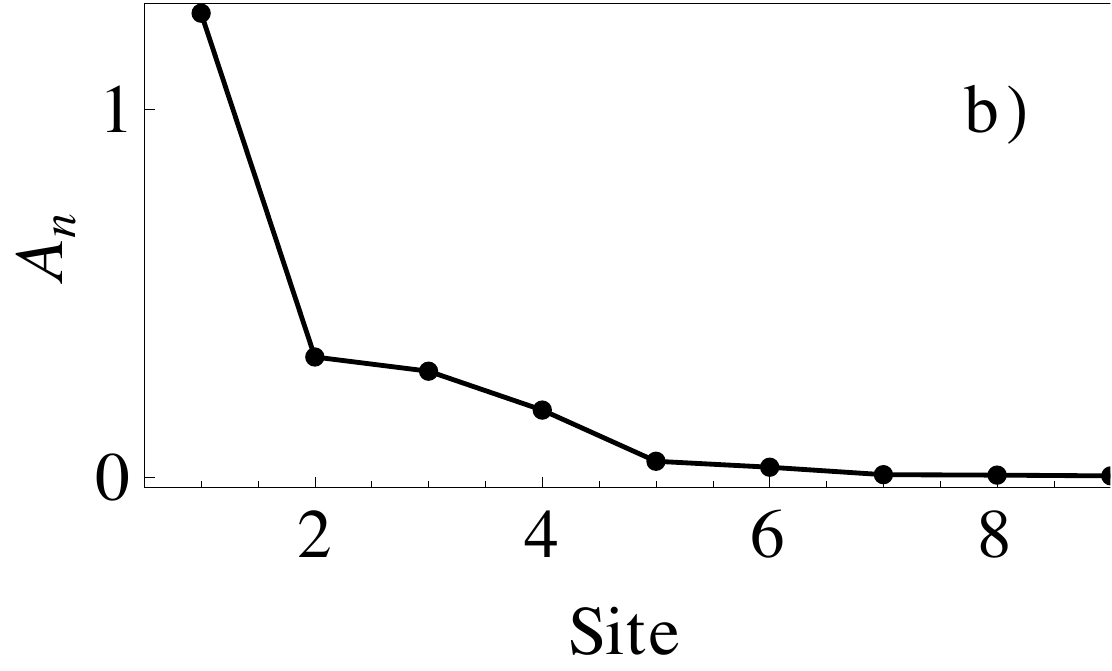}
\includegraphics[width=3.5cm]{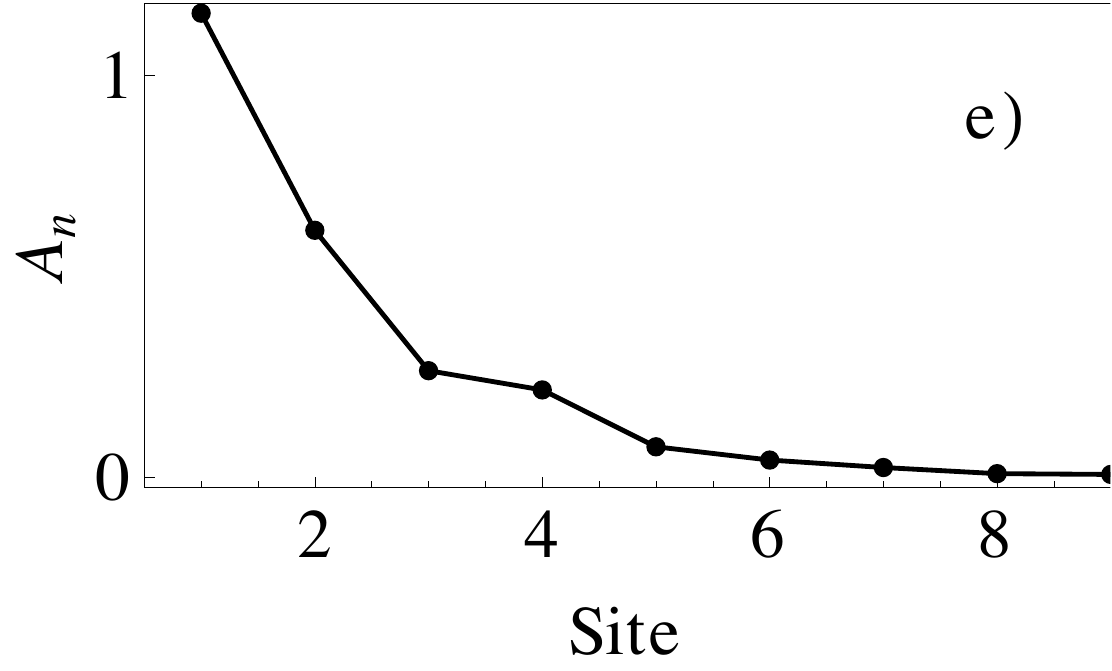}\\
\includegraphics[width=3.5cm]{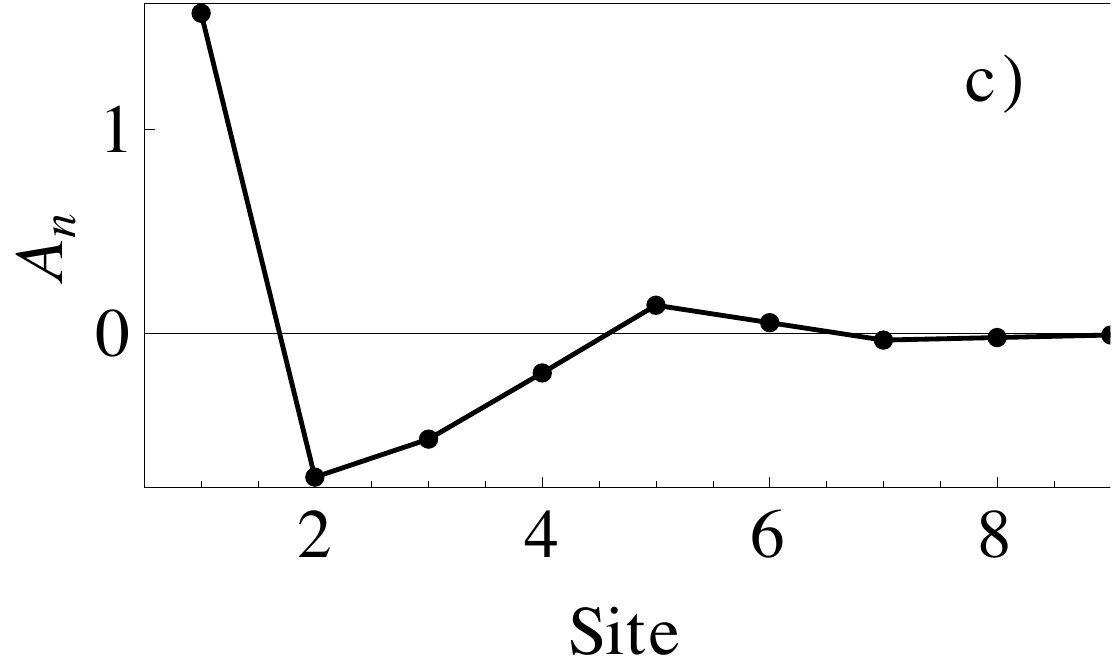}
\includegraphics[width=3.5cm]{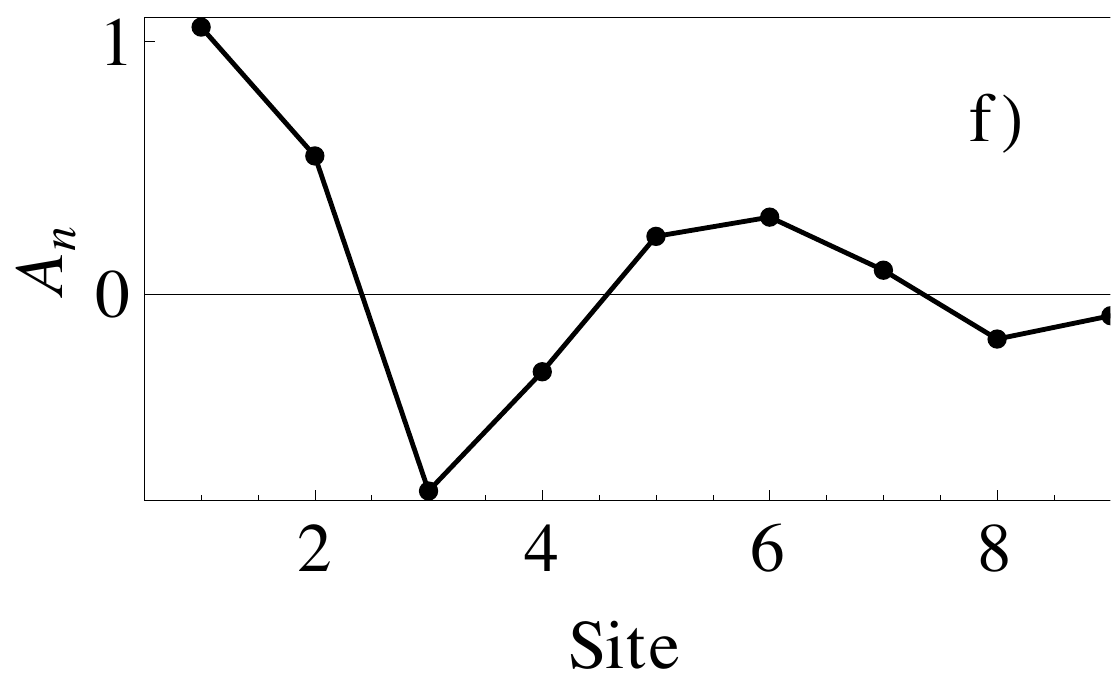}
\caption{Stable nonlinear surface modes for values of power and propagation constants marked
in Fig.\ref{fig2}. Left: Fibonacci model. Right: AA model.}
\label{fig3}
\end{figure}
Figure \ref{fig2} shows power content $P$ vs propagation constant $\lambda$ curves for different families of 
nonlinear surface modes that inhabit the different gaps of the Fibonacci and AA models, as well as their stability, for both the focussing and defocussing regime. We observe
that these modes originate from two main mechanisms: (i) Tangent bifurcations, usually
observed in the external gaps, and giving rise to one stable and one unstable mode. Here, the local environment is predominant and the actual size of the array does not play a significant role.
(ii) There are also nonlinear surface modes that bifurcate from linear surface modes (Fig.1a),
or from extended asymmetric modes (Fig.1b). The most interesting feature of
Fig.\ref{fig2} is the strong asymmetry between the focussing and defocussing 
cases, for the same distribution $\{\epsilon_{n}\}$ distribution.
 
In Fig. \ref{fig3} we show some examples of
surface modes for parameter values in different regions and marked in Fig.\ref{fig2}. They look qualitatively similar in both, the Fibonacci and AA models. Modes resembling these have also been observed in semi-infinite binary nonlinear waveguide 
arrays~\cite{Molina-binary}.

\subsection{Total Internal Reflection (TIR) modes}

\begin{figure}[t]
\includegraphics[width=6cm]{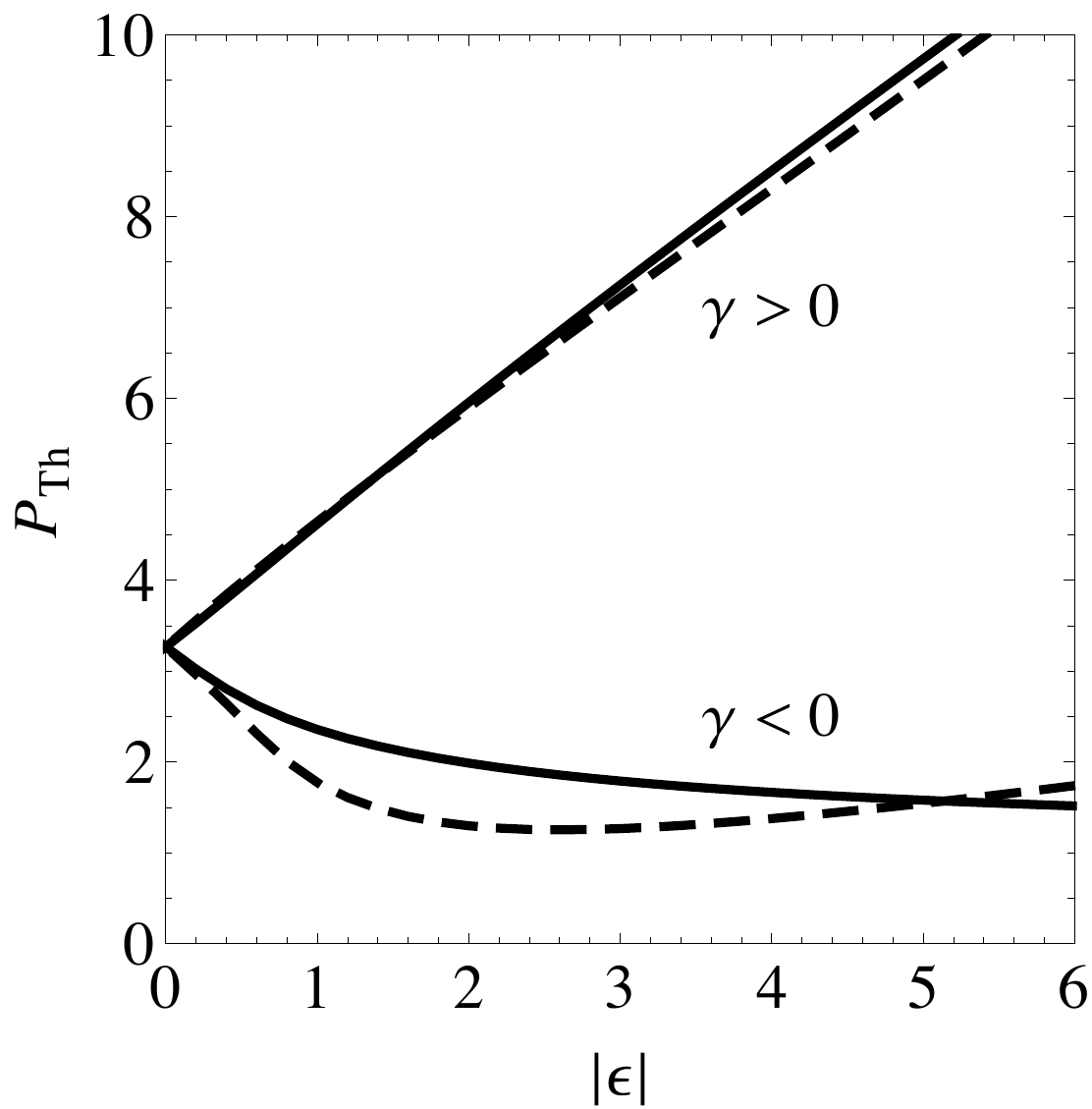}
\caption{Power threshold for excitation of TIR modes as a function of $\epsilon>0$
for focussing and defocussing regimes. Solid (dashed) curves correspond to the Fibonacci 
(Aubry-Andr\'{e}) models. $\gamma>0$ and $\gamma<0$ are associated to
Type I and II nonlinear modes, respectively.}
\label{fig4}
\end{figure}

TIR modes are highly-localized surface modes whose propagation constant resides
in the outermost external gap, where total internal reflection mechanisms 
dominate.
Usually, they need a threshold power to be excited. We observe that, for Type I modes,
this minimum power increases with an increase in $|\epsilon|$ (Fig.\ref{fig4}) in both 
models. For Type II modes, however, the behavior is different for each model. While for
the Fibonacci model, the power threshold decreases monotonically with an increase in
$|\epsilon|$, for the AA model, the threshold decreases initially, reaching a minimum
at $|\epsilon|\approx 2.59$, where the minimum threshold is
$P_{th}\approx 1.25$, and increasing monotonically with $|\epsilon|$
afterwards (Fig.\ref{fig4}).

A linear stability analysis reveals that all these TIR modes undergo a stability change
when $\partial P/\partial\lambda=0$, where the bifurcation occurs. 
The stability of the modes seems
also to associated with a different position of its `center of mass': In stable modes (Fig.3a,b,d,e), the
center of mass is closer to the surface, while unstable modes have their center of mass about one 
layer below the surface.

\subsection{Bragg Reflection(BR) modes}

These modes reside in deeper internal gaps in-between mini-bands, where Bragg reflection 
effects predominate. In these regions, localization is the combined result
of nonlinearity plus the scattering produced by the aperiodic structure.
These modes are less localized than the TIR modes (see Fig.3c,f), and they do not necessarily
need a power threshold to excite them. Modes with no threshold bifurcate from 
two types of linear modes: Those localized at the surface and associated to isolated
points of the linear spectrum (Fig1a), or those extended asymmetric modes where its envelope
decreases away from the surface (Fig1b). In any case, the presence of nonlinearity shifts the propagation constant of the
linear mode outside the mini-band, giving rise to a nonlinear surface mode. 

In the linear Fibonacci case, and for a relatively large number of sites there are a 
finite number of linear modes localized at the surface. This number is 
a function of 
the total number of sites, $N$. In particular, for $N=89$ there is a single localized
linear surface mode (Fig1a), which gives rise to the families of Type
I and II shown in Fig.2. On the other hand, for the Aubry-Andr\'{e} case, all of the linear eigenfunctions are extended for $|\epsilon|<2$ and localized for $|\epsilon|>2$\cite{Morandotti-AA}, 
and some of them are asymmetric
with maxima at or near the surface(Fig.1b). After the localization 
transition, these modes do not necessarily become localized 
at the surface (Fig.1c), but there are other linear modes that
will become localized at the surface. When nonlinearity is present, we note that 
solitons residing in the internal gaps,
between internal mini-bands, suffer a kind of oscillatory instability when their propagation
constant approaches the edge of a mini-band, and their power content increases. This is due
to a resonance between a linear mode at the edge of the band and the tail of the nonlinear mode, leading to a delocalization of the nonlinear mode and to a synchronization of its phase
with the one of the linear mode.
Similar behavior has been reported in
continuous systems with periodic potentials~\cite{11}, and also in
discrete binary systems~\cite{Molina-binary,6}.

\begin{figure}
\includegraphics[height=4.2cm]{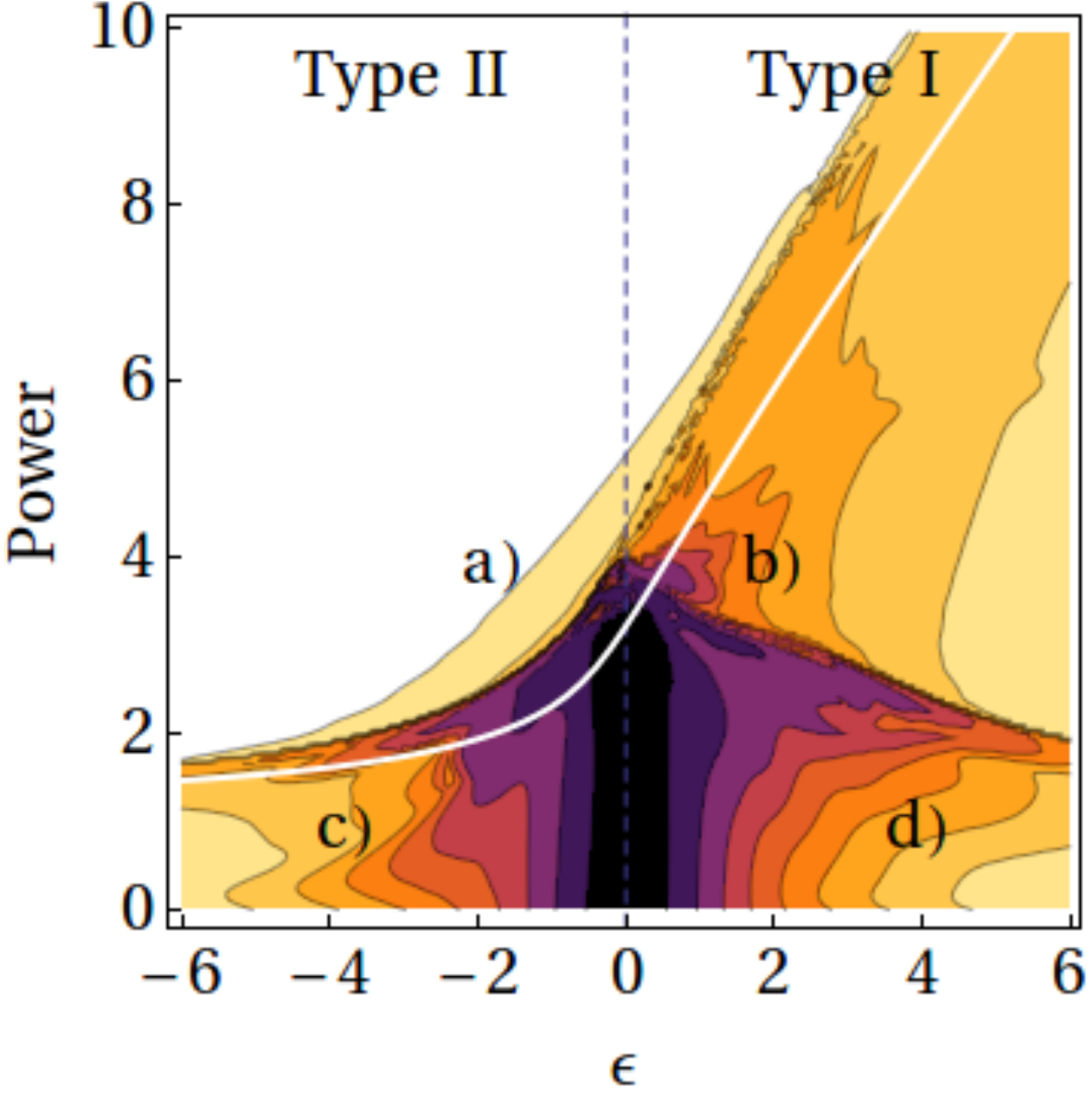}
\includegraphics[height=4.2cm]{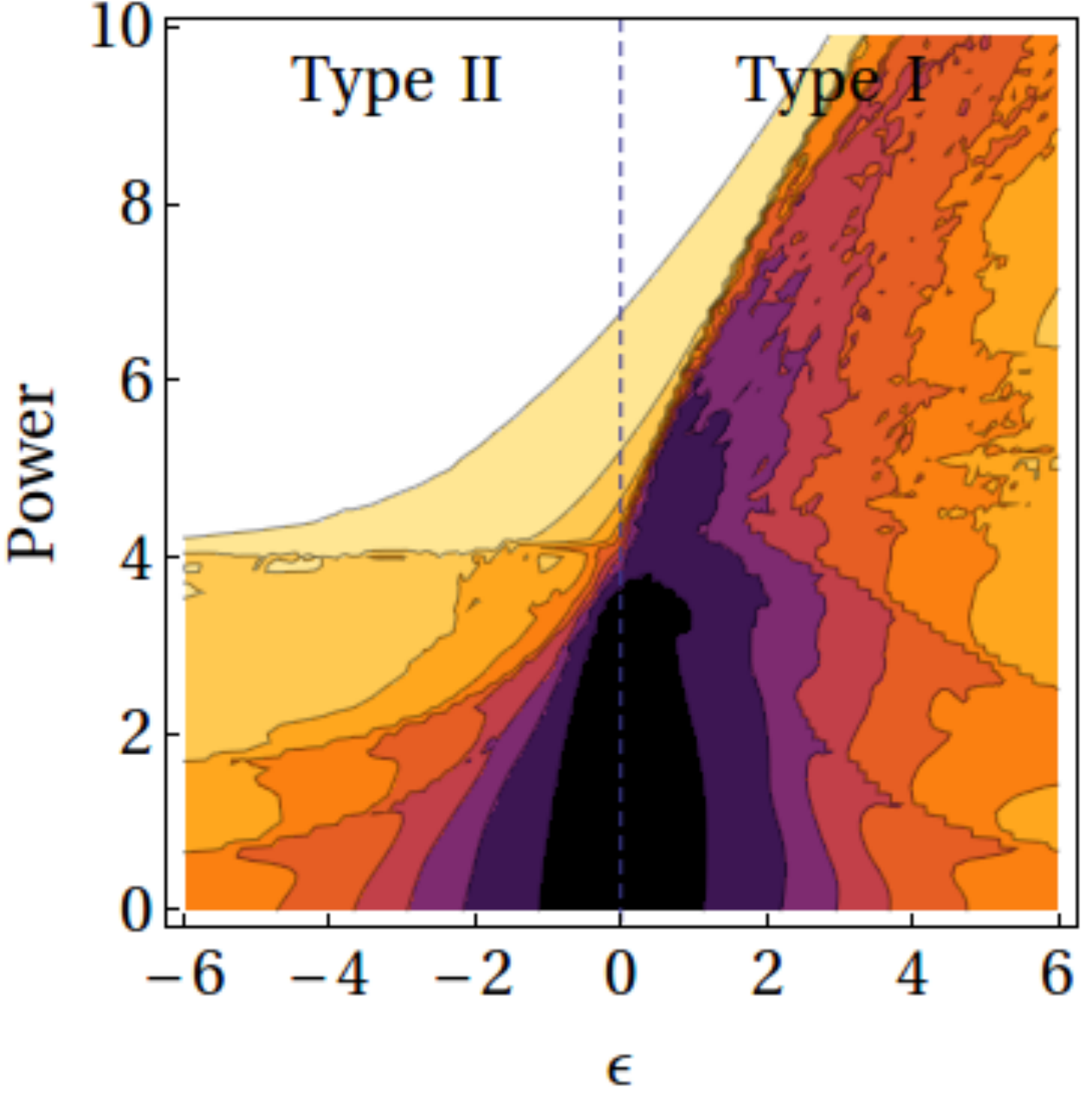}\\
\includegraphics[height=4.2cm]{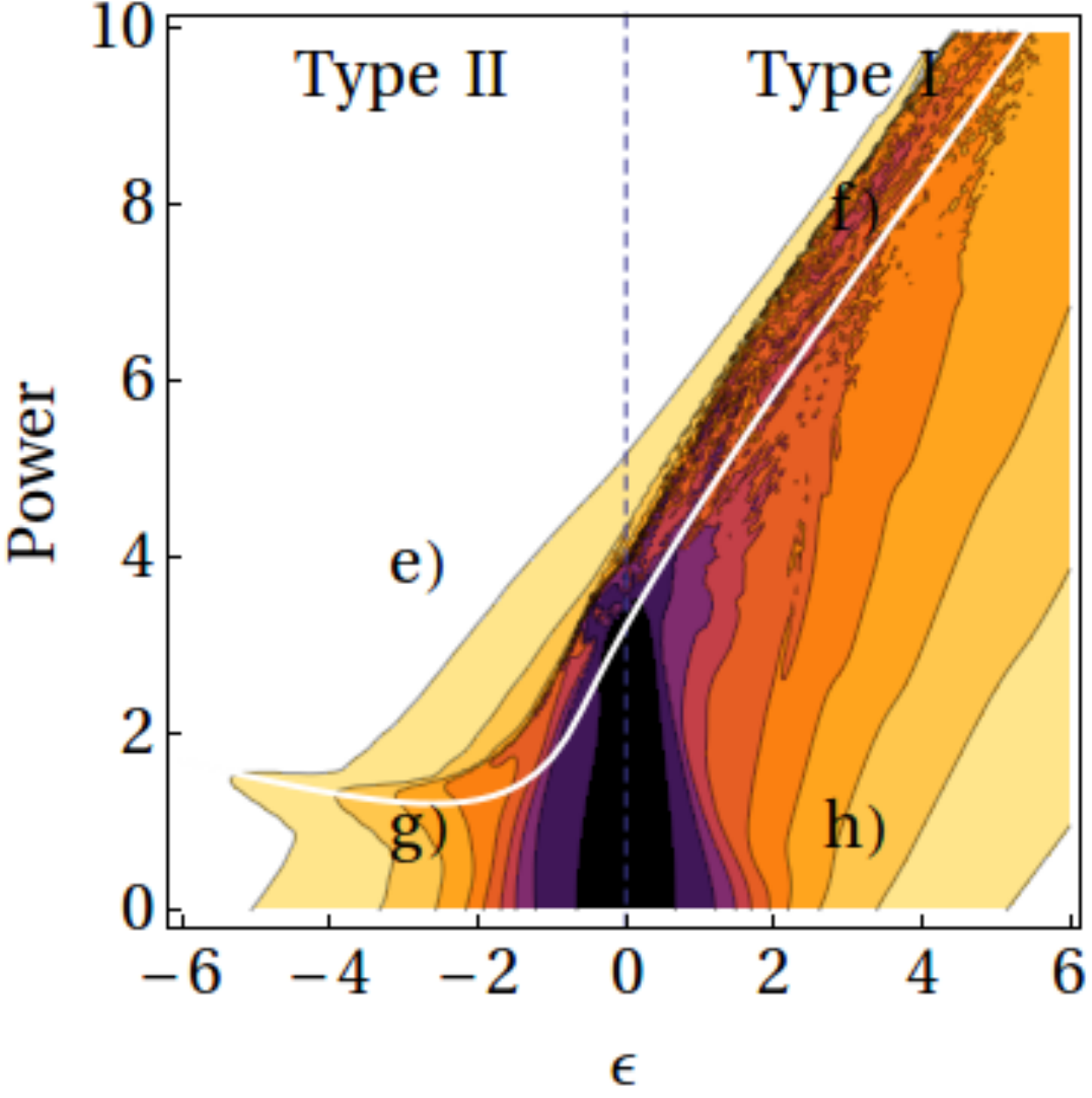}
\includegraphics[height=4.2cm]{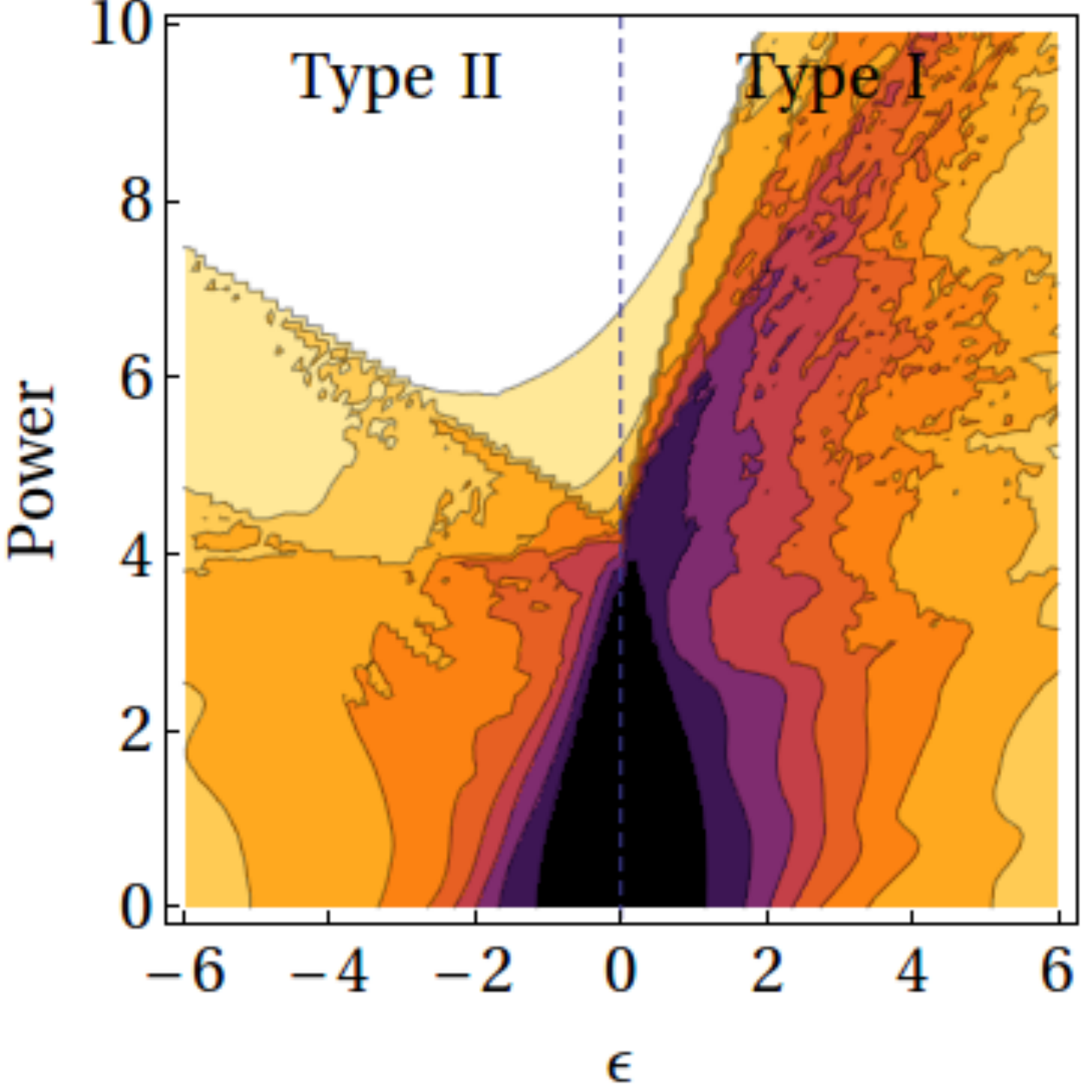}
\caption{(Color online). Output power fraction at initial site vs quasiperiodic strength
after a propagation  distance of $\xi_{max}=20$. First (second) row:
results for Fibonacci (Aubry-Andr\'{e}) arrays and, left (right) columns:
results for edge (bulk) excitation. Light and dark regions
denote high and low degree of localization. The white
curve denotes the power threshold to excite a TIR stationary surface mode
shown in Fig.\ref{fig4} (only for surface cases). The dynamics
of the marked states is shown in Fig.\ref{fig6}. In bulk cases we averaged over
3 realizations with different initial excite sites.}
\label{fig5}
\end{figure}

\section{DYNAMICS}

We look now into the effects of quasiperiodicity on the dynamical evolution of an 
initially localized input beam at the edge of the array and also 
inside the bulk, for comparison purposes. We solve Eq.(\ref{eq:2})
numerically with an initial condition of $A_n(0)=\sqrt{P}\ \delta_{n_0,n}$, 
with $n_0=1$ for the edge excitation. For the bulk case, we have different 
local `environments' around the initial site $n_0$ depending 
on the specific chosen site, we average the selftrapped fraction obtained for three different initial sites. 

For a given $\epsilon$ and $P$, the space-averaged fraction of power
remaining at the initial waveguide $f$, after a longitudinal 
propagation distance $\xi_{max}$ is defined as 
$f=(P\ \xi_{max})^{-1}\int_0^{\xi_{max}}|A_1(\xi)|^2d\xi$. 
Results for $f$ as a function of $\epsilon$ and $P$ with $\gamma>0$ 
for both models, are shown in Fig.\ref{fig5} in the
form of a density plot. 

\begin{figure*}[t]
\includegraphics[height=3.5cm]{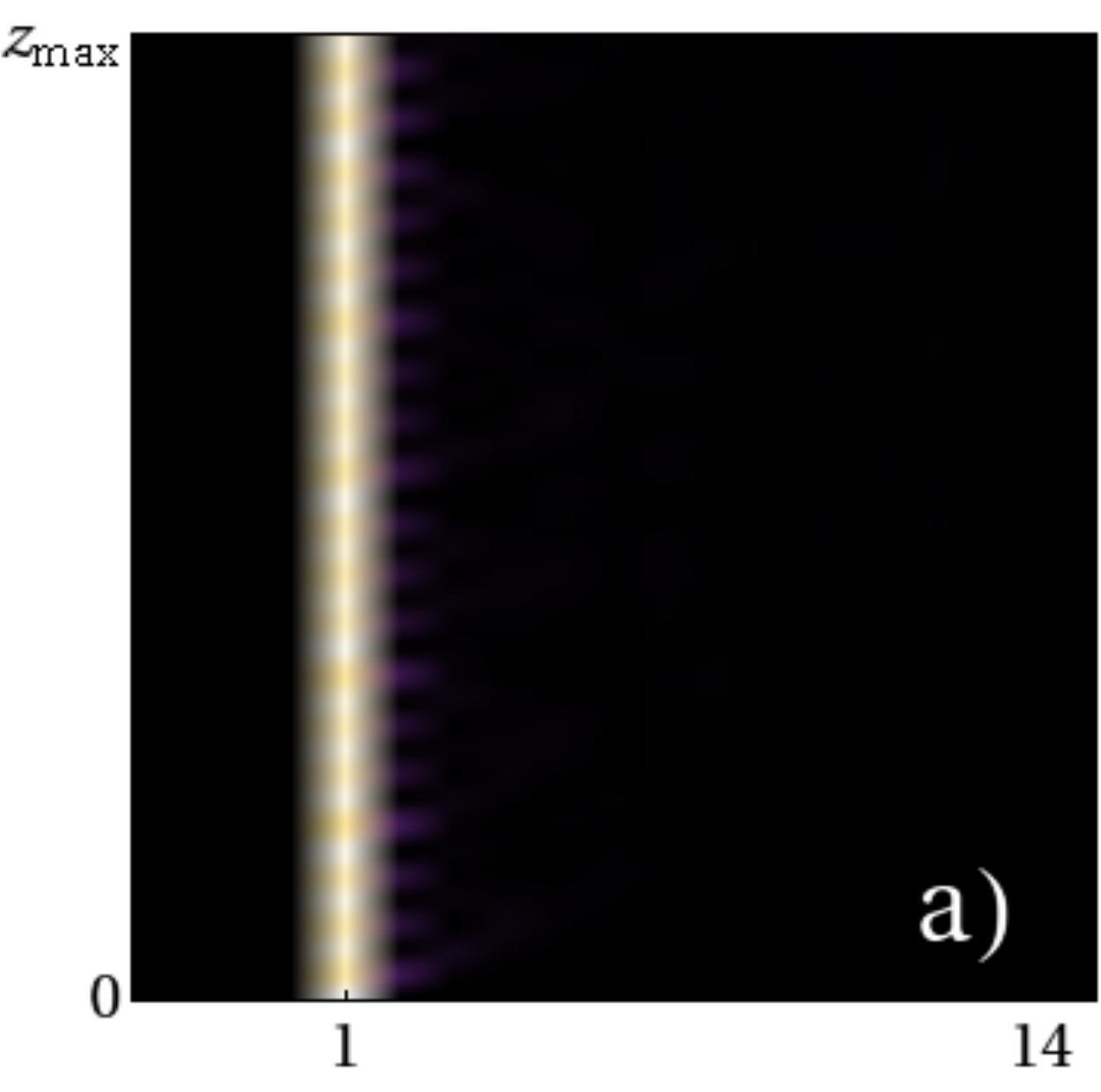}
\includegraphics[height=3.5cm]{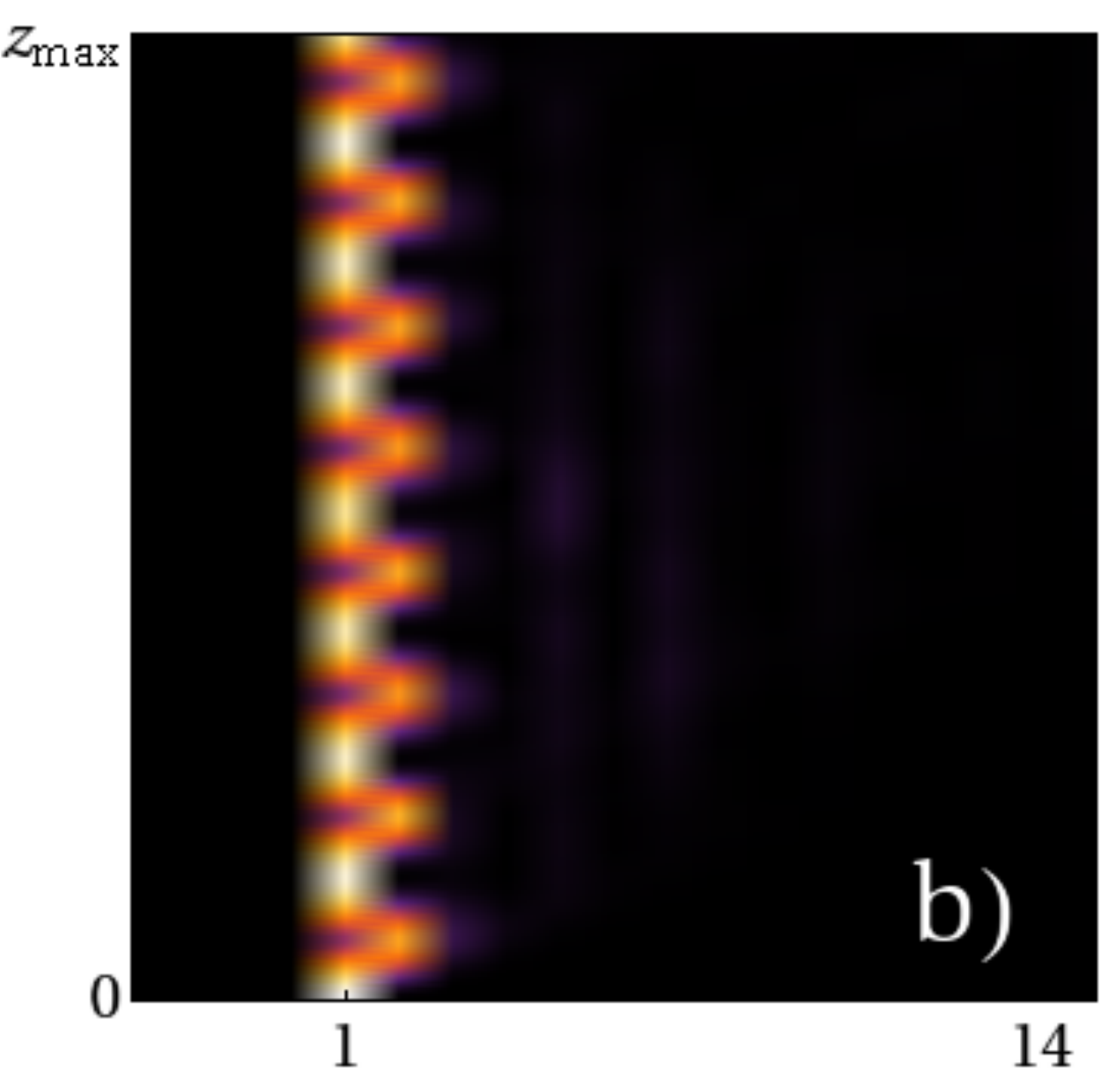}
\includegraphics[height=3.5cm]{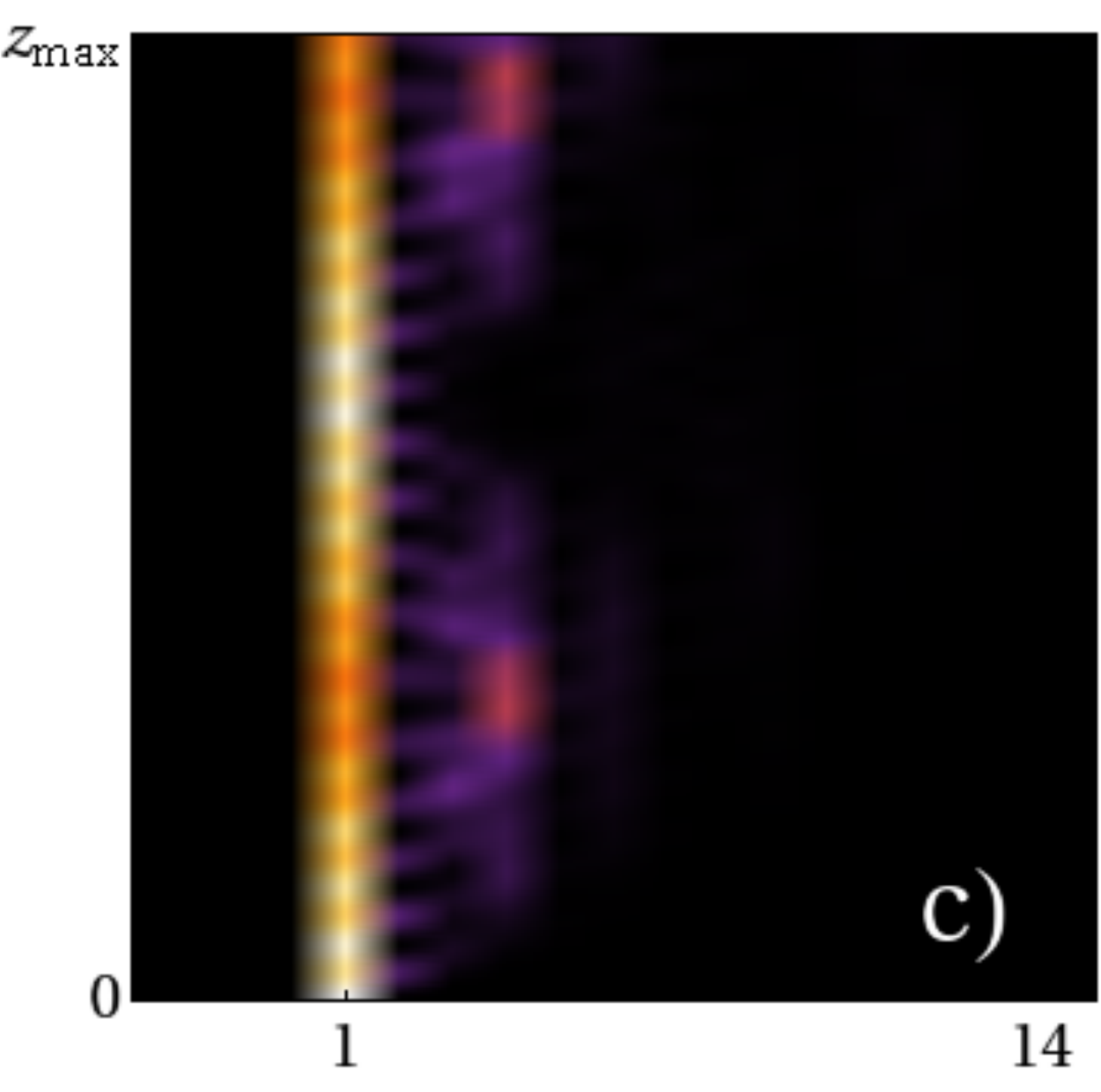}
\includegraphics[height=3.5cm]{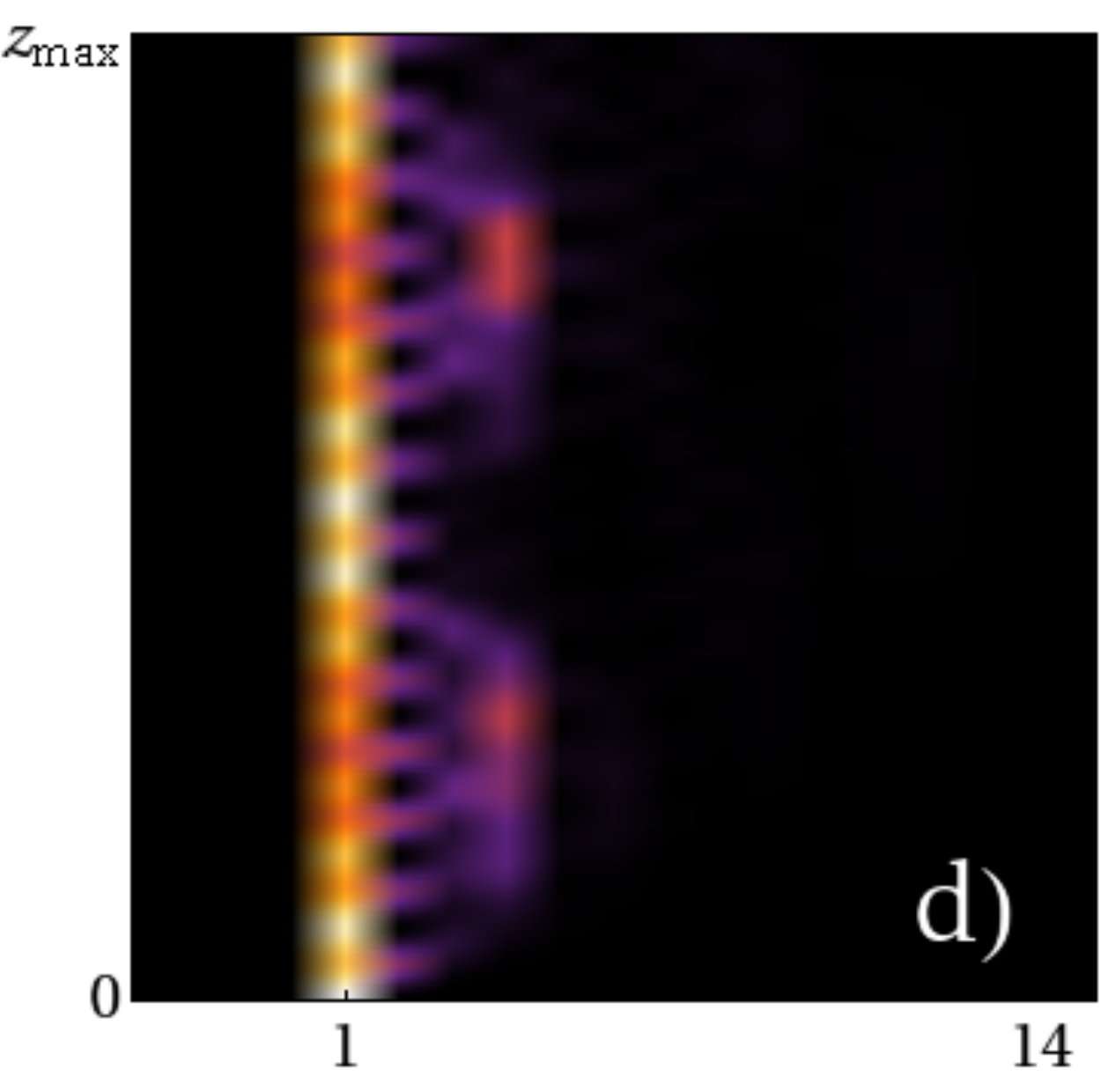}\\
\includegraphics[height=3.5cm]{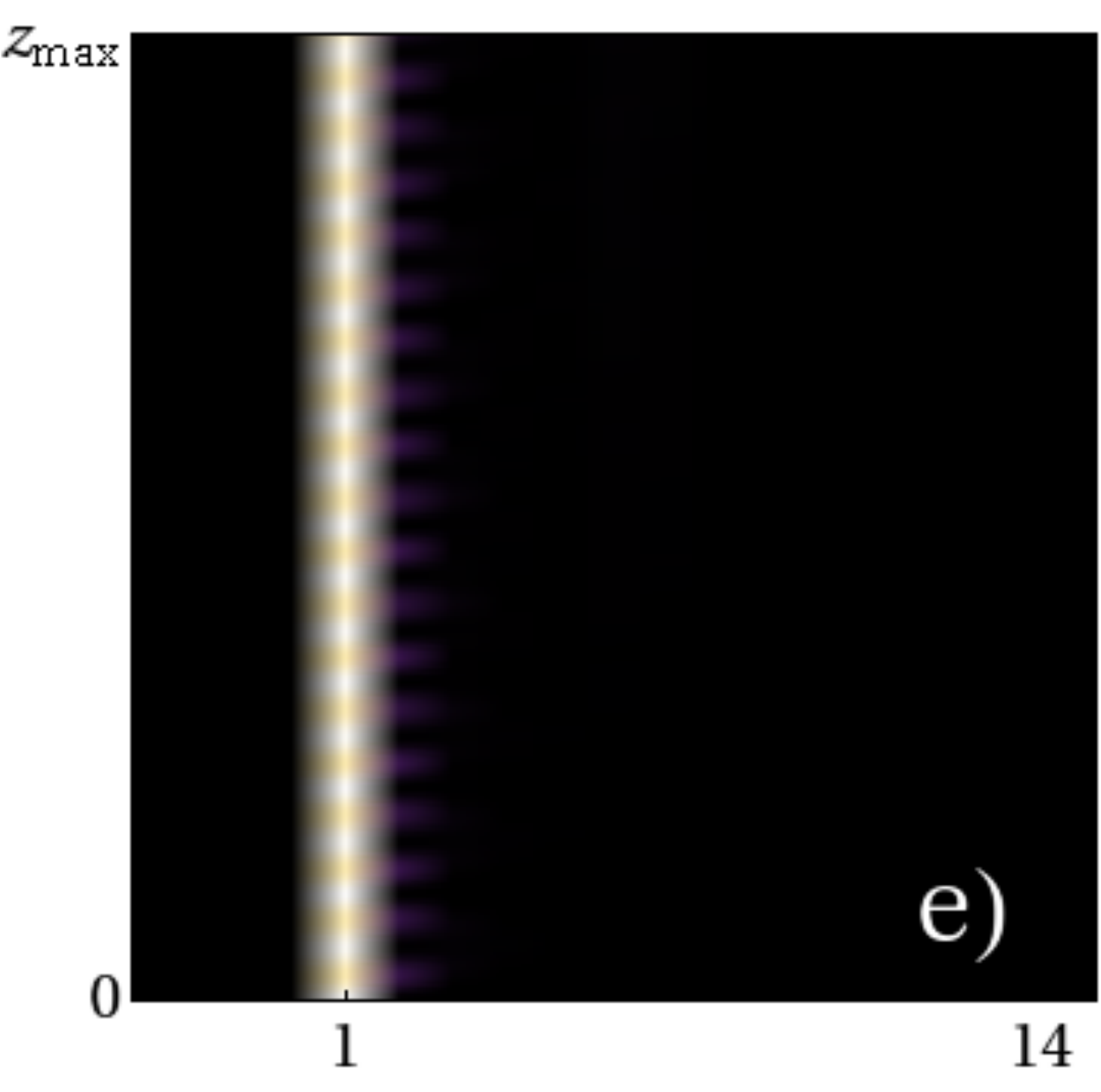}
\includegraphics[height=3.5cm]{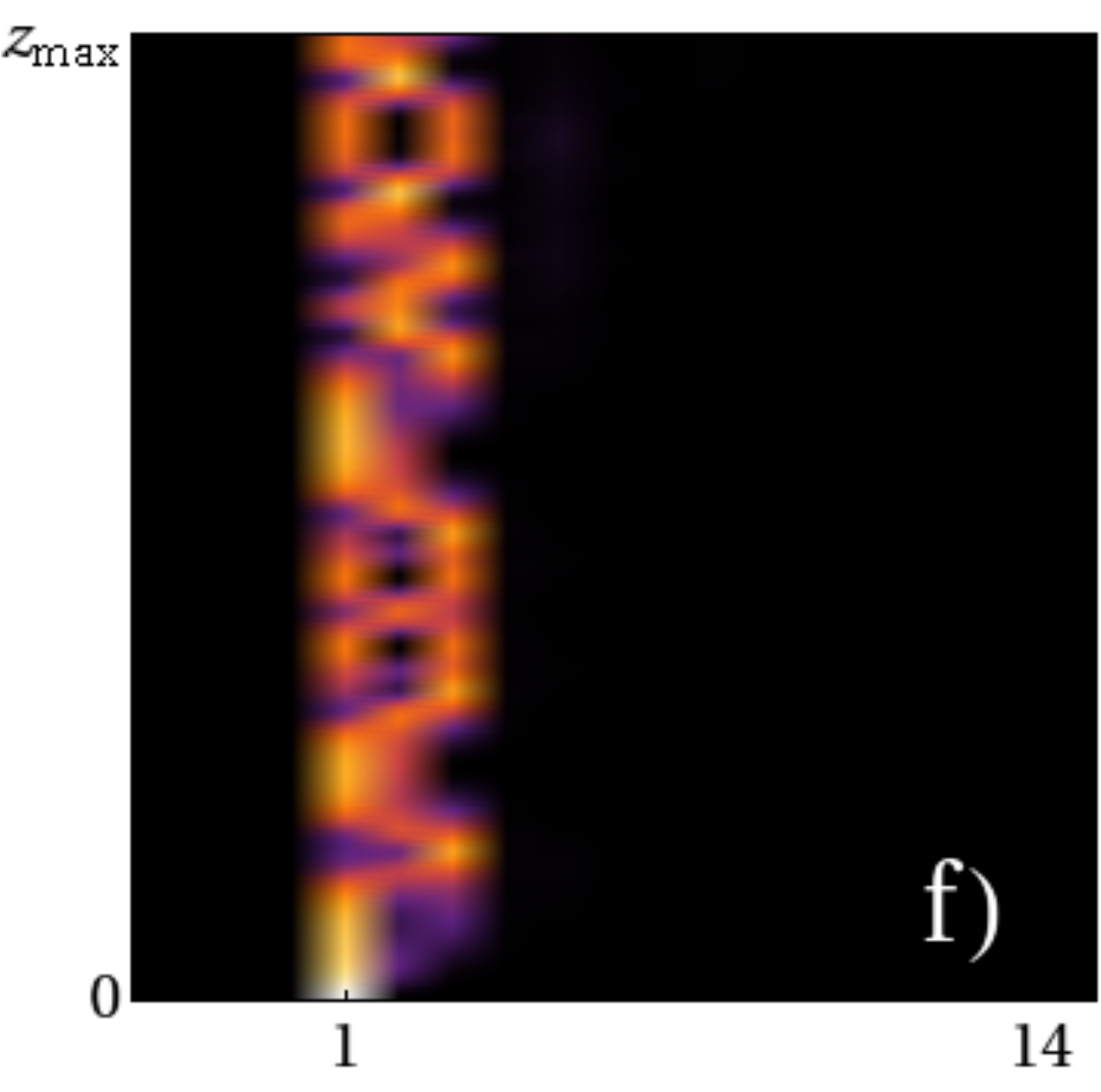}
\includegraphics[height=3.5cm]{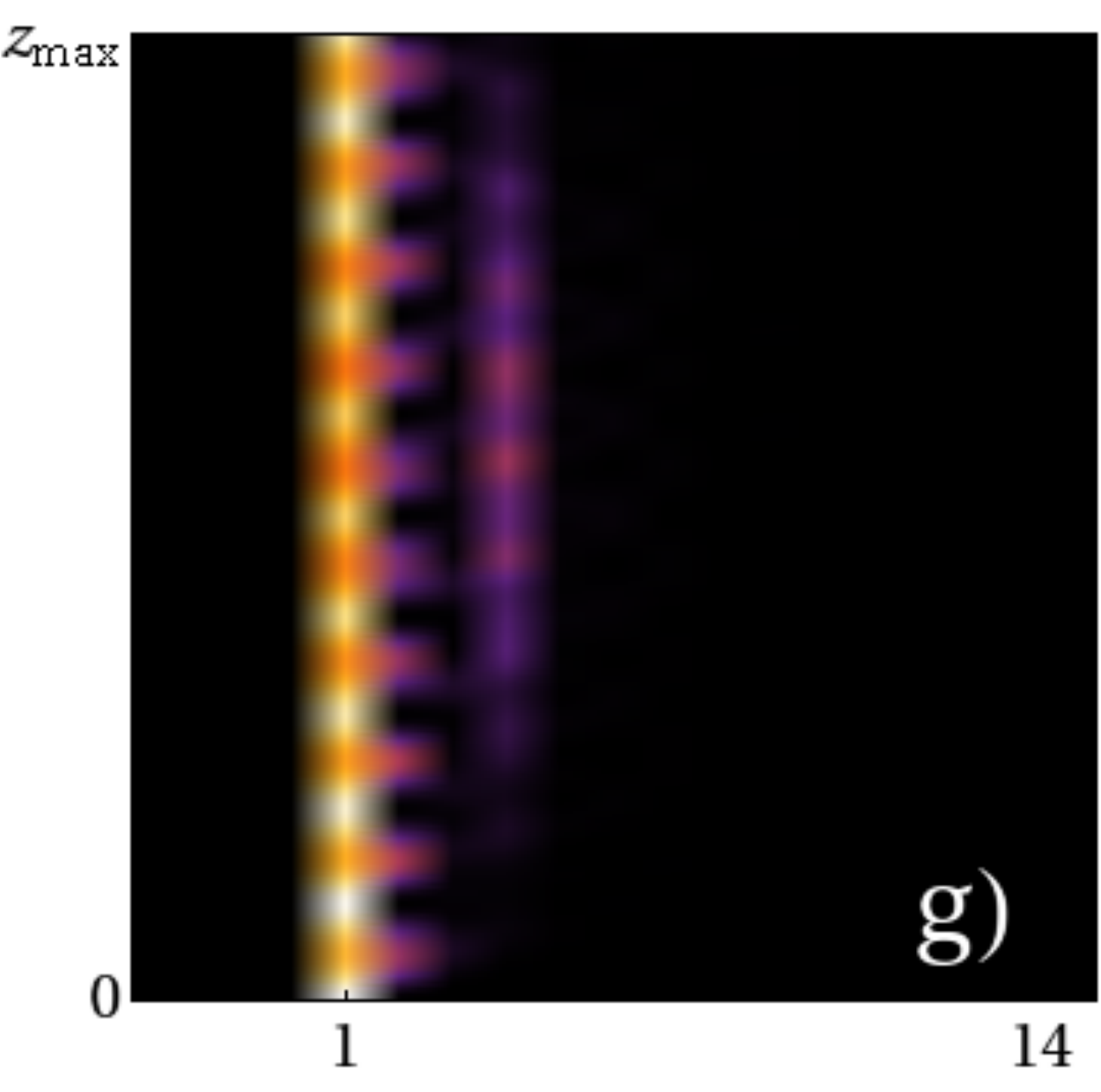}
\includegraphics[height=3.5cm]{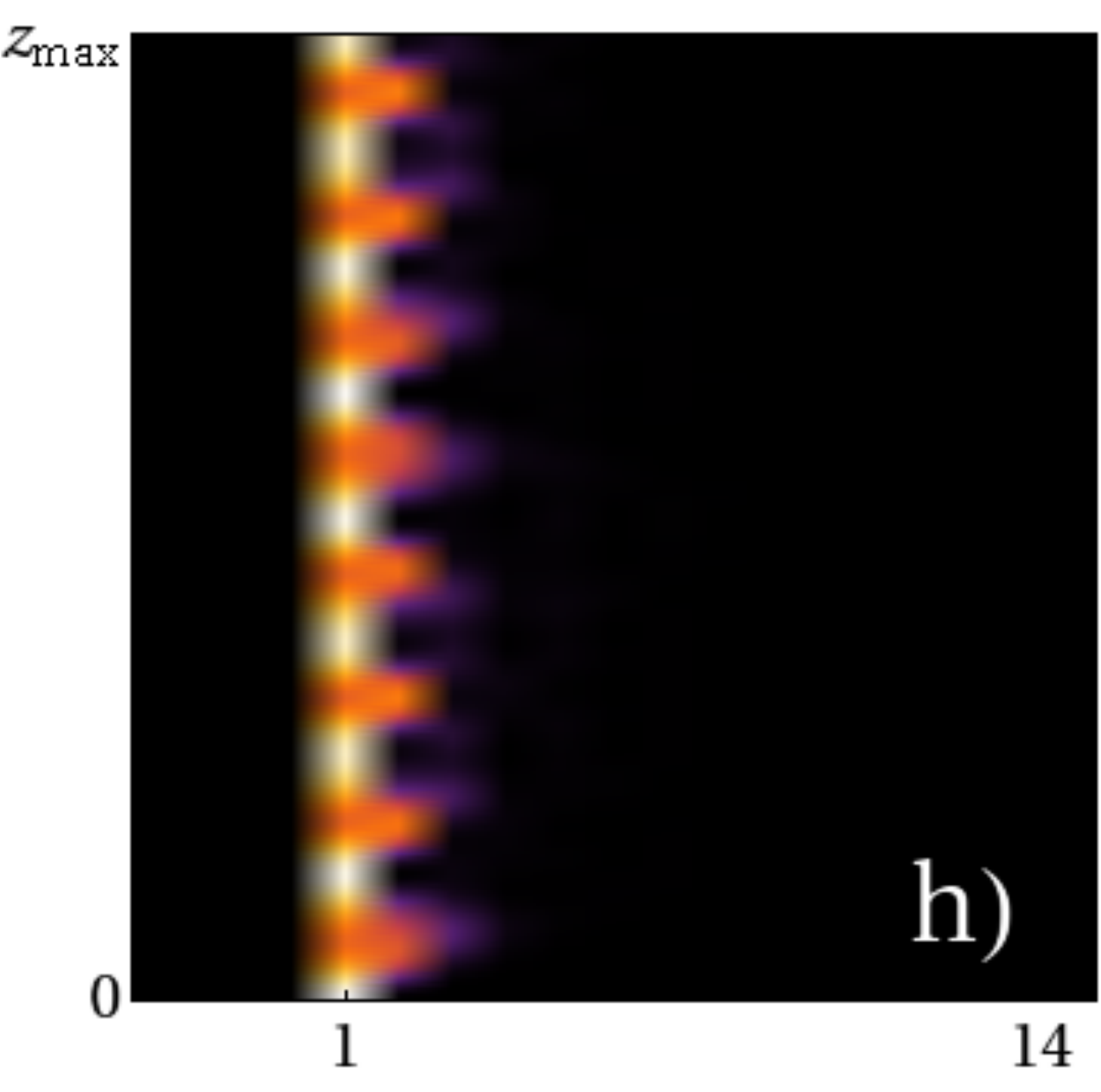}
\caption{(Color online). Dynamical evolution of a delta-like initial input beam, in the form of a density plot showing the amplitude $|A_n(z)|^2$ as a function of longitudinal distance $z$
and waveguide position $n$. Plots (a)-(d) and (e)-(f) correspond to the
Fibonacci and AA models, respectively. ($z_{max}=20, N=89)$.}
\label{fig6}
\end{figure*}

The first thing we notice from Fig.\ref{fig5} is that both, Fibonacci and AA models, show a qualitative similar selftrapped fraction as a function of the
aperiodic strength. The most interesting feature, however, is that
the threshold power to excite localized modes is actually {\em lower} at the surface than at the bulk. More specifically, we note that
for a given power content, a smaller quasiperiodic strength  
is required to effect localization at the surface than in the bulk. And for fixed quasiperiodic strength, a 
smaller power is needed to localize the excitation at the edge than inside the bulk. This is in marked contrast with the behavior observed in periodic systems.

Coming back to the surface excitation case, we note a correspondence between the threshold power 
to excite a discrete surface soliton dynamically, with the thresholds for
the stationary modes (white line) of the previous section. The dynamical thresholds 
are always larger. It is also interesting the marked asymmetry between the region
to the right of $\epsilon=0$ (Type I nonlinear modes) with the region to the left 
of $\epsilon$ (equivalent to Type II nonlinear modes). This 
asymmetry mirrors the one
found for stationary surface modes. As expected, in the region where nonlinearity dominates,
the output power fraction at the surface guide is close to unity, while at low input powers
and high quasiperiodic strength, there is a sort of weak localization near the surface. 
At very small input power, the beam tends to diffract below certain 
quasiperiodic strength. Conversely, when
quasiperiodicity is very weak, the beam diffracts below a certain 
minimum input power. Different localization regimes are separated by
regions where it is not possible to localize the wave function. 
Eventually, when the power or the strength of the
quasiperiodic potential ($\epsilon$) is sufficiently high, most of
the optical power remains trapped. 

Some examples of the detailed dynamical evolution in different regions
of Fig.\ref{fig5} for surface excitations, is shown in Fig.\ref{fig6}. We see different types of localized surface solutions:\\
(a) High-power solutions that are strongly localized at the surface and are 
associated to the TIR gap (Fig.6a and 6e). Here, localization increases with power.\\
(b) Surface solutions that oscillate persistently between two sites, with a 
propagation constant that depends on the value of $\epsilon$ (Fig.6b and 6h).
We have computed the Fourier transform of these solutions 
and found that the propagation constant associated to the transversal dynamics 
(secondary propagation constant) is smaller than the propagation
constant associated to the
longitudinal dynamic (primary propagation constant), which dominates the dynamics. An increase in 
$|\epsilon|$ render these type of solutions more localized and the
secondary propagation constant becomes negligible. However, their power content is 
smaller than the power associated with the TIR modes. This 
type of solution is similar 
to the ones described in Ref.\cite{7} for a lattice with alternating site
energies, where stable dynamic exists with the power oscillating
between two sites.\\
(c) Low-power solitons that show oscillations with two different
propagation constants (Figs. 6c, 6d and 6g): The first oscillation is fast and occurs between two nearby 
sites (as in case (b)); the second oscillation is slower and involves sites farther away from the surface (about $5$ sites).\\
(d) Chaotic oscillations (Fig. 6f), where the oscillations do not seem to follow a 
well-defined pattern. These solitons reside at the very boundary between weak and strong selftrapped output fraction in the $P_0-\epsilon$ diagram in Fig.\ref{fig5}

\section{NONDIAGONAL QUASIPERIODICITY}

For completeness, we also report here the results obtained for a variant of the Fibonacci model, where
the aperiodic sequence occurs for the coupling between nearest-neighbor guides,
and the individual refractive index contrast of the guides are taken as identical. In that case, it can be  absorbed as a phase factor in the solution, which we take equal to zero, without loss of generality. The equations of motion are:
\begin{equation}
i\frac{d}{d\xi}A_n + V_{n+1} A_{n+1}+V_n A_{n-1}
+\gamma|A_n|^2A_n
= 0
\end{equation}
where $V_{n}$ denotes the coupling between guides $n-1$ and $n$. According to the Fibonacci substitutional rule, The $\{V_{n}\}$ sequence is $V_a\ V_b\ V_a\ V_a\ V_b\ 
V_a\ V_b{\ldots}$

The results obtained (not shown here) are qualitatively similar to the previous case
of quasiperiodicity in the refractive index 
(diagonal quasiperiodicty), with a linear spectrum whose structure is composed of bands and gaps, with each band in turn,
decomposed into smaller mini-bands and gaps. As in the previous case, nonlinear surface modes of the TIR and BR type can also be excited.  
For this diagonal quasiperiodic case however, the usual staggered-unstaggered symmetry holds:
$(\lambda, \gamma,A_{n})\rightarrow(-\lambda,-\gamma,(-1)^n A_{n})$
as in the usual DNLS equation. This implies that the nonlinear surface modes for the focussing and defocussing cases are now equivalent.

\section{DISCUSSION}

We have examined the interplay among nonlinearity, quasiperiodicity
and surface effects in a semi-infinite, one-dimensional array of weakly-coupled, nonlinear (Kerr) optical waveguides. We have considered two implementations of quasiperiodicity, namely, the Fibonacci and the 
Aubry-Andr\'{e} model, and have found different families of surface localized modes, as well as the dynamical evolution of an initial surface input beam. Both models  show qualitatively similar behavior in the nonlinear regime, characterized by high-power solitons (TIR gap) strongly localized at the edge of the array, and lower-power solitons (BR gap) that oscillate persistently between some few sites close to the edge. Perhaps the main result is the strong asymmetry 
observed between the focussing and defocussing cases for the same
distribution $\{\epsilon_n\}$, that could be measured 
experimentally using techniques similar to the ones used in ref~\cite{Morandotti-AA}. Thus, for a nonlinear focussing medium, the defocussing case could also be explored by a judicious choice of the aperiodic strength parameter $\epsilon$. For instance, in the Aubry-Andr\'{e} model,
where the modulation of the lattice is given by $\eta\cos(2\pi n\xi)$, if the choice
$\{\eta>0,\gamma>0\}$ gives rise to focussing modes, then to explore defocussing ones, is enough to change $\eta\rightarrow-\eta$, which in turn, is equivalent to explore the modes
in the defocussing regimen $\{\eta>0,\gamma<0\}$. 
A comparison of the dynamical evolution of excitations initially localized at the surface and at the bulk, revealed that for a fixed degree of quasiperiodic strength, a smaller amount of optical power is required to selftrap the excitation at the surface than at inside the bulk. Conversely, for a fixed optical power, a larger quasiperiodic strength is required to selftrap the excitation inside the bulk than at the surface. To summarize, it is easier to localize an excitation at the edge than inside the bulk. This is in marked contrast with the behavior observed for periodic photonic lattices, and cast a note of caution on the widespread notion that a one-dimensional surface acts in a `repulsive' manner.

\begin{acknowledgements}
The authors are grateful to R. Vicencio, M. Johansson and Y. S. Kivshar for useful discussions. This work was supported in part by FONDECYT Grants 1080374 and
1070897, and Programa de Financiamiento Basal de CONICYT (Grant FB0824/2008).
\end{acknowledgements}


\begin{thebibliography}{}

\bibitem{AL-exp}
T. Pertsch et. al., Phys. Rev. Lett. 93, 053901 (2004);
T. Schwartz, G. Bartal, S. Fishman and M. Segev, {\em Nature} {\bf 446}, 52 (2007);
Yoav Lahini, Assaf Avidan, Francesca Pozzi, Marc Sorel, Roberto Morandotti, Demetrios Christodoulides and Yaron Silberberg, Phys. Rev. Lett. 100, 013906 (2008).

\bibitem{tamm}
C. R. Rosberg, D. N. Neshev, W. Krolikowski, A. Mitchell, R. A. Vicencio, M. I. Molina and
Yu. S. Kivshar, Phys. Rev. Lett. {\bf 97}, 083901 (2006);
E. Smirnov et al., Opt. Lett. 31, 2338 (2006). 

\bibitem{Szameit-anderson}A. Szameit, Y. V. Kartashov, P. Zeil, F. Dreisow, M. Heinrich,
R. Keil, S. Nolte, A. T\"unnerman, V. A. Vysloukh and L. Torner, Opt.
Lett. {\bf 35}, 1172 (2010).

\bibitem{nanopillars}
S. V. Zhukovsky, D. N. Chigrin, and J. Kroha, J. Opt. Soc. Am B {\bf 23},
2265 (2006);

\bibitem{metal-fibonacci}
Xiao-Ning Pang, Jian-Wen Dong, and He-Zhou Wang, J. Opt. Soc. Am {\bf 27}, 2009 (2010).




\bibitem{Morandotti-AA}
Y. Lahini, R. Pugatch, F. Pozzi, M. Sorel, R. Morandotti, N.
Davidson and Y. Silberberg, Phys. Rev. Lett. {\bf 103}, 013901 (2010).

\bibitem{AA}
S. Aubry and G. Andr\'{e}, Ann. Isr. Phys. Soc. {\bf 3}, 133 (1980).

\bibitem{faso}E. S. Zijstra, A. Fasolino and T. Janssen, Phys. Rev. B {\bf
59}, 302 (1999).

\bibitem{niu}Qian Niu and Franco Nori, Phys. Rev. B {\bf 42},
10329 (1990).

\bibitem{Molina-Surface}
K. G. Makris et al., Opt. Lett. 30, 2466 (2005); 
S. Suntsov et al., Phys. Rev. Lett. 96, 063901 (2006);
M. I. Molina, R. A. Vicencio and Yu. S. Kivshar, Opt. Lett. {\bf 31}, 1693 (2006); .

\bibitem{MLT}
M. I. Molina, N, Lazarides and G. P. Tsironis, unpublished.

\bibitem{ILM}
D. K. Cambpell, S. Flach, and Y. S. Kivshar, {\em Phys. Today} {\bf 57}, 43 (2004).

\bibitem{molecular crystals}
A. A. Ovchinikov, Sov. Phys. JETP 30, {\bf 147} (1970); R. Bruinsma,
K. Maki, and J. Wheatley, Phys. Rev. Lett. {\bf 57}, 1773 (1986); A. S. Davydov, {\em Solitons in Molecular Systems}, translated by E. S. Kryachko (Kluwer, Dordrecht, 1985).

\bibitem{biopolymers}
A. Xie, L. van der Meer, W. Hoff, and R. H. Austin, Phys. Rev. Lett. {\bf 84}, 5435 (2000); T. Dauxois and M. Peyrard, in {\em Nonlinear Excitations in Biomolecules}, edited by M. Peyrard (Springer, New York, 1995), p. 127; J. C. Eilbeck, P. S. Lomdahl, and A. C. Scott, Physica D {\bf 16}, 318 (1985); A. C. Scott, Phys. Rep. {\bf 217}, 1 (1992); A. Scott, {\em Nonlinear Science: Emergence and Dynamics of Coherent Structures}, 2nd ed. (Oxford University Press, New York, 2003); G. P. Tsironis, M. Iba–es, and J. M. Sancho, Euro- phys. Lett. {\bf 57}, 697 (2002); S. F. Mingaleev, Y. B. Gaididei, P. L. Christiansen, and Y. S. Kivshar, ibid. {\bf 59}, 403 (2002).

\bibitem{junctions}
L. M. Flor\'{\i}a, J. L. Mar\'{\i}n, P. J. Mart\'{\i}nez, F. Falo, and S. Aubry, Europhys. Lett. {\bf 36}, 539 (1996); E. Trias, J. J. Mazo, and T. P. Orlando, Phys. Rev. Lett. {\bf 84}, 741 (2000); P. Binder, D. Abraimov, A. V. Ustinov, S. Flach, and Y. Zolotaryuk, ibid. {\bf 84}, 745 (2000); A. Ustinov, Chaos {\bf 13}, 716 (2003).

\bibitem{BE}
A. Trombettoni and A. Smerzi, Phys. Rev. Lett. {\bf 86}, 2353 (2001); A. Trombettoni and A. Smerzi, J. Phys. B {\bf 34}, 4711 (2001); A. Smerzi, A. Trombettoni, P. G. Kevrekidis, and A. R. Bishop, Phys. Rev. Lett. {\bf 89}, 170402 (2002).

\bibitem{basics}
M. Johansson, Ph. D. Thesis, Link\"{o}ping University, Sweden, 1995. 

\bibitem{AA_theory}
J. Avron, Simon B, Duke Math. J. {\bf 50}, 369 (1983);
Y. Last, Commun. Math. Phys. {\bf 164}, 421 (1994).


\bibitem{Molina-binary}Mario I. Molina, Ivan L. Garanovich, Andrey A. Sukhorukov and
Yuri S. Kivshar, Opt. Lett. {\bf 31}, 2332 (2006).


\bibitem{11}
A. A. Sukhorukov and Yu. S. Kivshar, Opt. Lett. 27 2112 (2002); 28, 2345 (2003); Phys. Rev. Lett. 91, 113902 (2003); A. A. Sukhorukov,
Yu. S. Kivshar, H. S. Eisenberg and Y. Silberberg, IEEE J. Quantum Electron. 39, 31 (2003).


\bibitem{6}
Rodrigo A. Vicencio and Magnus Johansson, Phys. Rev. A {\bf 79}, 065801 (2009).

\bibitem{7}
M. Johansson and A. V. Gorbach, Phys. Rev. E 70, 057604 (2004).















\end{thebibliography}
\end{document}